\documentclass[aps,preprintnumbers,amsmath,amssymb,preprint,eqsecnum,floatfix]{revtex4}

\usepackage{graphicx}
\usepackage{euscript}
\usepackage{oldgerm}
\usepackage{subfigure}

\begin{document}

\title{Superposition of Quantum and Classical Rotational Motions in Sc$_\text{2}$C$_\text{2}$@C$_\text{84}$
Fullerite}
\date{\today}
\author{K.H. Michel$^1$, B. Verberck$^1$, M. Hulman$^{2,3}$, H. Kuzmany$^2$, and
M. Krause$^{2,4}$}
\affiliation{$^1$Departement Fysica, Universiteit Antwerpen, Groenenborgerlaan 171, 2020 Antwerpen, Belgium \\
$^2$Institut f\"{u}r Materialphysik der Universit\"{a}t Wien, Wien, Austria \\
$^3$ARC Seibersdorf Research GmbH, A-2444 Seibersdorf, Austria \\
$^4$Institute for Ion Beam Physics and Materials Research, Forschungszentrum
Rossendorf, PF510119, D-01314, Dresden, Germany}

\begin{abstract}
The superposition of the quantum rotational motion (tunneling) of the encapsulated Sc$_\text{2}$C$_\text{2}$
complex with the classical rotational motion
% (uniaxial diffusion and stochastic jump reorientations)
of the
surrounding C$_\text{84}$ molecule in a powder crystal of
Sc$_\text{2}$C$_\text{2}$@C$_\text{84}$ fullerite is investigated by theory.
Since the quantum rotor is dragged along by the C$_\text{84}$ molecule,
any detection method which couples to the quantum rotor (in casu the C$_\text{2}$ bond of the
Sc$_\text{2}$C$_\text{2}$ complex) also probes
the thermally excited classical motion (uniaxial rotational diffusion and stochastic
meroaxial jumps) of the surrounding fullerene.
The dynamic rotation-rotation response
functions in frequency space are obtained as convolutions of quantum and classical dynamic correlation functions.  The
corresponding Raman scattering laws are derived, the overall shape of the spectra and the width of the resonance
lines are studied as functions of temperature.  The results of the theory are confronted with experimental
low-frequency Raman spectra on powder crystals of Sc$_\text{2}$C$_\text{2}$@C$_\text{84}$ [M.\ Krause et al.,
Phys.\ Rev.\ Lett.\ {\bf 93}, 137403 (2004)].  The agreement of theory with experiment is very satisfactory in a
broad temperature range.
\end{abstract}

\maketitle

%---------------------
\section{Introduction}
%---------------------
The existence of an endohedral fullerene, i.e.\ one or several atoms encapsulated in a fullerene molecule, was originally
inferred from an analysis of mass spectra of LaCl$_\text{3}$-impregnated graphite and did lead to the proposal
La@C$_\text{60}$ \cite{1}.  At present the study of endohedral metallofullerenes M$_x$@C$_n$, $x=1,2,3,4$ and
$n=66, 68, 72, 74,\hdots, 100$, where M are group II and III metals such as Sc, Y, $\hdots$ or lanthanides
Ce, $\hdots$, Lu, is a subject of interdisciplinary research \cite{2} in physics, chemistry and materials
sciences.  By now one is able to produce materials where not only single atoms but clusters of atoms are
encapsulated \cite{3}. 
Due to charge transfer between the cluster and the surrounding carbon cage it is possible to obtain molecular-like
complexes which do not exist otherwise (i.e.\ in absence of encapsulation) and which have unusual properties.
Not only clusters with metal atoms of
a same kind, such as the Sc$_\text{3}$ trimer in Sc$_\text{3}$@C$_\text{82}$ are produced, but also clusters
composed of different kinds of atoms.  A remarkable case is the production of
Sc$_\text{2}$C$_\text{2}$@C$_\text{84}$ in crystalline powder form \cite{4}.
The powder crystal is composed of crystallites where the Sc$_\text{2}$C$_\text{2}$@C$_\text{84}$ units are
arranged with average space
group symmetry $Fm\overline{3}m$.  From spectroscopic and structural
characterization by NMR- and synchrotron X-ray diffraction experiments \cite{4} it follows that the
Sc$_\text{2}$C$_\text{2}$ complex is encaged as a rigid unit with point group symmetry $D_{2h}$ in a
C$_\text{84}$ fullerene of symmetry D$_{2d}$ (isomer III, number 23 \cite{5}).  The center of mass of
Sc$_\text{2}$C$_\text{2}$ coincides with the center of mass of the molecule.
The two Sc atoms are located at a distance of $4.29(2)$ {\AA} on the long
$C_\text{2}$ ($S_4$) axis of C$_\text{84}$.  
The two C atoms of Sc$_\text{2}$C$_\text{2}$ are located in the plane containing the two $C_\text{2}$ axes
perpendicular to the long axis of the C$_\text{84}$ molecule and
have a calculated distance of $1.28$ {\AA}.  This distance lies between
those of typical double and triple carbon bonds, and is consistent with the experimental and calculated C--C
stretching frequency of the C$_\text{2}$ unit (exp.\ $1745$ cm$^{-1}$, calc.\ $1742$ cm$^{-1}$) \cite{6}.  In the following we will
speak of this C--C bond as a
C$_\text{2}$ unit or molecule.  Indeed low energy Raman spectra \cite{6} on powder
samples of Sc$_\text{2}$C$_\text{2}$@C$_\text{84}$ in a temperature range $25$ -- $150$ K (Kelvin)
have revealed the existence of quantized rotational states of the C$_\text{2}$ unit.
%These states result from rotations of the Sc$_\text{2}$C$_\text{2}$ complex about its long axis which coincides
%with the long axis of the C$_\text{84}$ molecule.
The Raman lines' positions
reflect
% are in good agreement with transitions between
transitions between energy levels of a C$_\text{2}$ planar quantum rotor in a fourfold static potential due to
the surrounding C$_\text{84}$ cage.  
Therefore, one can speak of a quantum gyroscope.

In Ref. \onlinecite{6} the potential parameters of the encaged quantum rotor
were obtained from density functional calculations using the
VASP (Vienna ab initio simulation package) code \cite{9}.  The energy levels were then determined from the solution of the Schr\"{o}dinger
equation.  Within this approach, the Raman spectra consist of infinitely
sharp lines while experimentally the lines are
broadened and have a characteristic temperature behavior.  A reason for this shortcoming is the restriction of
the role of the encapsulating C$_\text{84}$ molecule to a purely static body.  
Since the measured transition frequencies are in the range of $10$ -- $80$ cm$^{-1}$ and since the line broadening
is of the order of a few cm$^{-1}$, any involvement of internal vibrational modes
of the C$_\text{84}$ cage as well as of stretching modes Sc--C$_\text{84}$
can be excluded.  The latter are of
higher frequencies and have been measured in C$_\text{84}$ and in Sc$_\text{2}$@C$_\text{84}$ by infrared and
Raman techniques \cite{Krause2,Krause3}.  However the low-frequency external rotational modes of
the C$_\text{84}$ molecule and their superposition with the transitions of the quantum rotor should be retained:
indeed the encapsulated Sc$_\text{2}$C$_\text{2}$ gyroscope is
dragged by
the classical rotational motion of the C$_\text{84}$ molecule and this dragging will affect the Raman spectrum of the
C$_\text{2}$ unit.  It
follows that an experimental probe such as Raman scattering which couples to the encapsulated species in an
endohedral complex, in casu Sc$_\text{2}$C$_\text{2}$, also yields information on the dynamics of the
encapsulating molecule, in casu C$_\text{84}$.

In the present paper we will extend the theoretical interpretation given in Ref. \onlinecite{6} and develop a
unified theory where the quantum
mechanical motion of the Sc$_\text{2}$C$_\text{2}$ complex
is coupled to the thermally excited classical rotational
motion of the C$_\text{84}$ fullerene.  
The coupling results from the fact that the long axis of the quantum gyroscope coincides with the $S_4$ axis of
the surrounding C$_\text{84}$ molecule.
In a given crystallite the C$_\text{84}$ molecules are randomly
oriented with their long $C_2$ axis in equivalent $\langle 100 \rangle$ directions of the face-centered cubic (fcc)
unit
cell \cite{4}.  We call this situation meroaxial disorder (this terminology seems to be more appropriate than merohedral
disorder).
We start from a model where at low temperature the meroaxially oriented
C$_\text{84}$ molecules in the fcc crystal perform uniaxial rotational diffusions about their long axis. 
% In reality the C$_\text{84}$
% molecules in the fcc crystal perform uniaxial rotations about their long axis.
Such a classical motion can be seen as a
time-dependent modulation of the fourfold potential experienced by the quantum rotor and causes a
temperature-dependent broadening of the quantum levels.
An additional broadening effect is to be expected from the stochastic reorientations of the
C$_\text{84}$ molecules among the meroaxial directions which should become increasingly
important at higher temperature.  In
addition, the stochastic reorientations lead to the appearance of a temperature-dependent
quasi-elastic peak in the Raman spectrum.

The content of the paper is as follows.  In Section II we write down the Raman scattering law for the
C$_\text{2}$-unit belonging to the Sc$_\text{2}$C$_\text{2}$ complex encapsulated by the C$_\text{84}$ molecule
in Sc$_\text{2}$C$_\text{2}$@C$_\text{84}$ fullerite.  We start from a single crystal with $Fm\overline{3}m$
structure and static meroaxial disorder of the C$_\text{84}$ molecules.  Assuming a quantum mechanical
rotational motion of the Sc$_\text{2}$C$_\text{2}$ complexes and a classical rotational motion of the
C$_\text{84}$ molecules, the dynamic polarizability-polarizability correlation function is decoupled in a
product of correlation functions for the rotational dynamics of Sc$_\text{2}$C$_\text{2}$ and C$_\text{84}$
respectively.  The scattering law is obtained as a convolution of these correlation functions
in Fourier space.
Next (Sect.\ III) we calculate the correlation functions, using a quantum mechanical tunneling model for the 
Sc$_\text{2}$C$_\text{2}$ complex and a uniaxial rotational diffusion model for the surrounding C$_\text{84}$
molecule.  The rotational diffusion motion of the encapsulating C$_\text{84}$ molecule leads to a linear
temperature-dependent broadening of the energy
transition lines of the C$_\text{2}$ planar rotor.  In Sect.\ IV we extend the theory to a powder crystal
which consists of arbitrarily oriented crystallites, each with $Fm\overline{3}m$ space group symmetry and static
meroaxial disorder.  In the following (Sect.\ V) we consider the case of dynamic
meroaxial disorder, describing the reorientations of C$_\text{84}$ molecules among the three meroaxial
directions by a stochastic jump model.
This model yields an exponential temperature-dependent broadening of the transition lines.
In the last Section VI we give a numerical evaluation of the Raman
scattering law for a Sc$_\text{2}$C$_\text{2}$@C$_\text{84}$ powder crystal where quantum mechanical tunneling
of the encapsulated Sc$_\text{2}$C$_\text{2}$ units is superimposed by uniaxial rotational diffusion and dynamic
meroaxial disorder of the
C$_\text{84}$ molecules.  The temperature dependence of the line intensities and of the line broadenings is
discussed.

\section{Raman scattering law}\label{secRaman}
We will derive the Raman scattering law where we limit ourselves to the interaction of the incident laser light
with the plane rotational motion of the induced dipole of the C--C bond belonging to the Sc$_\text{2}$C$_\text{2}$ complex of
Sc$_\text{2}$C$_\text{2}$@C$_\text{84}$.
%We consider a model where Sc$_\text{2}$C$_\text{2}$ is taken as a rigid
%unit of $D_{2h}$ symmetry.  The surrounding C$_\text{84}$ molecule has $D_{2d}$ symmetry \cite{3} (isomer
%III).
This means that we consider the low frequency part of the spectrum (say $\le 100$ cm$^{-1}$).
% The high-frequency part which accounts for the vibrational modes has been investigated in Ref.\ \onlinecite{Krause2}.
The Sc$_\text{2}$C$_\text{2}$ complex is centered in the origin (center-of-mass position) of C$_\text{84}$.
The long axis of Sc$_\text{2}$C$_\text{2}$ coincides with the $S_4$ axis of C$_\text{84}$.  The C--C bond of
Sc$_\text{2}$C$_\text{2}$ lies in the plane containing the secondary $C_2$ axes
of C$_\text{84}$ and rotates about the $S_4$ axis. 
In that respect we will consider the C--C bond as a C$_\text{2}$ planar rotor which experiences a fourfold
potential inside the C$_\text{84}$ molecule.
Our formulation of the Raman scattering law is an extension of the conventional theory \cite{10,11} in as much as
we describe a situation where the rotational motion of the induced dipole with respect to the laboratory-fixed
frame is a superposition of the quantum motion of the C$_\text{2}$ planar rotor inside the C$_\text{84}$ molecule
and of the classical motion of the C$_\text{84}$ molecule in the laboratory frame.

\begin{figure}
%\subfigure{
\resizebox{5cm}{!}
{\includegraphics{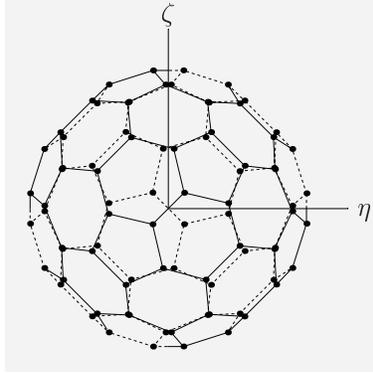}}
%}
%\subfigure{\resizebox{5cm}{!}
%{\includegraphics{c84zxtoepscnvingv.eps}}}
%\subfigure{\resizebox{5cm}{!}
%{\includegraphics{c84xytoepscnvingv.eps}}}
\caption{View of the C$_\text{84}$ molecule along the $S_4$ axis.}
\label{figC84}
\end{figure}

\begin{figure}
\resizebox{8cm}{!}{\includegraphics{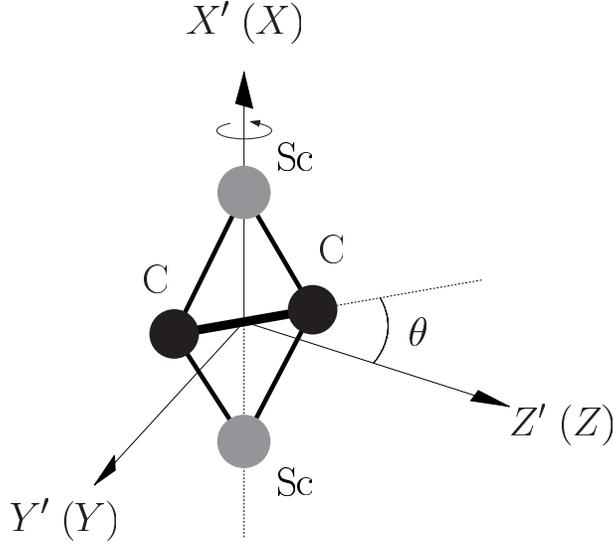}}
\caption{The Sc$_\text{2}$C$_\text{2}$ complex in the crystal-fixed cubic
coordinate system $(X',Y',Z')$ while the C$_\text{84}$ molecule is in standard
orientation.
}
\label{figaxes3D}
\end{figure}

\begin{figure}
\resizebox{8cm}{!}{\includegraphics{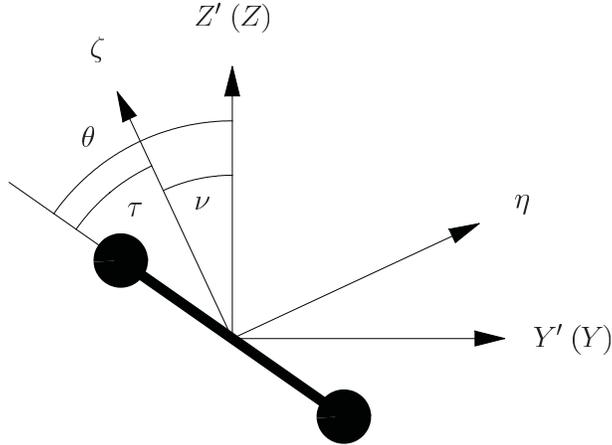}}
\caption{Orientation of the C$_\text{2}$ bond of Sc$_\text{2}$C$_\text{2}$ in
the rotatory reflection plane of the C$_\text{84}$ molecule (C$_\text{84}$ in
standard orientation).}
\label{figaxes2D}
\end{figure}

We start with considering a single crystal of Sc$_\text{2}$C$_\text{2}$@C$_\text{84}$ units with static meroaxial
disorder.  We assume that the Sc$_\text{2}$C$_\text{2}$@C$_\text{84}$ units are statistically independent, hence it
will be sufficient to consider one single representative unit.
The cubic crystal axes $(X',Y',Z')$ are chosen to coincide with the
laboratory-fixed cubic coordinate system $(X,Y,Z)$.  We consider a cubic system of axes $(\xi,\eta,\zeta)$ fixed in the
C$_\text{84}$ molecule such that the $\xi$ axis coincides with the $S_4$ axis while $\eta$ and $\zeta$ coincide
with the secondary twofold axes (Fig.\ \ref{figC84}).
The meroaxial orientations of the C$_\text{84}$ molecules correspond to the situation where the $\xi$ axes are
randomly oriented along the $X'$, $Y'$ or $Z'$ crystal axes.  The C$_\text{2}$ units then rotate in the planes
$(Y',Z')$, $(Z',X')$ or $(X',Y')$ respectively (Fig.\ \ref{figaxes3D}).
We say that the C$_\text{84}$ molecule is in
standard orientation if the $S_4$ axis coincides with the laboratory-fixed
$X$ axis and the plane containing the secondary $C_2$ axes coincides
with the laboratory
$(Y,Z)$ plane.  The $\zeta$ axis forms an angle $\nu$ with the $Z$ axis, while the C--C bond forms an
angle $\tau$ with the $\zeta$ axis.  Hence the polar angle $\theta$
of the C--C bond with the laboratory $Z$ axis (Fig.\ \ref{figaxes2D}) is a sum of two terms:
\begin{align}
   \theta = \nu + \tau. \label{angles}
\end{align}
Since the C$_\text{2}$ rotor is confined to the $(Y,Z)$ plane, the azimuthal angle $\phi$ measured away from $X$
has value $\pi/2$.  The distinction of two contributions to the angle $\theta$ is essential.

%---------
% Figure 2
%---------

%---------
% Figure 3
%---------

In the following we will assign the angular variable $\tau$ to the quantum mechanical tunneling of the
Sc$_\text{2}$C$_\text{2}$ complex about its long axis inside the C$_\text{84}$ cage and the angular variable
$\nu$
to the thermally excited classical rotation of the C$_\text{84}$ molecule about the $S_4$ axis.
The assumption of classical uniaxial rotational diffusion motion as a first approximation to the dynamics of the
C$_\text{84}$ molecule at low temperature is motivated by the structural results of meroaxial disorder \cite{4}. 
It is also inspired from the dynamics of solid C$_\text{70}$ in the rhombohedral and monoclinic
phases.  There the importance of uniaxial rotational diffusion about the long axis of the C$_\text{70}$ molecule
has been probed by muon spin spectroscopy \cite{Dennis,Mcrae}, nuclear magnetic resonance \cite{Maniwa,
Blinc, Tycko} and inelastic neutron scattering \cite{Christides}.

We treat the C--C bond of Sc$_\text{2}$C$_\text{2}$ as a rigid cylindrical rod with longitudinal and transverse
static polarizability $\alpha_\parallel$ and $\alpha_\perp$ respectively.
%  The C$_\text{2}^-$ rod rotates in the $YZ$ plane about the $X$ axis.
The Raman scattering law for incident
and scatterd radiation in $Z$ direction is given by the Fourier transform of the time-dependent autocorrelation
function of the polarizability $\alpha_{ZZ}$:
\begin{align}
   R_{ZZZZ}(\omega) = \frac{N}{2\pi}\int_{-\infty}^{+\infty} dt\,
   e^{i\omega t}\langle\alpha_{ZZ}(t)\alpha_{ZZ}(0)\rangle. \label{RZZZZ}
\end{align}
Here $N$ is the number of Sc$_\text{2}$C$_\text{2}$@C$_\text{84}$ units and $\omega$ is the frequency difference
of incident and scattered radiation.  The polarizability has to be
understood as an average over the three meroaxial molecular orientations.  In the following we label these
orientations by a superscript $(i)$, $i=1,2,3$.
If the C$_\text{84}$ molecule is in standard orientation ($\xi$ axis $\parallel X$),
or in orientation $\xi \parallel Y$, the corresponding orientation-dependent polarizabilities $\alpha_{ZZ}^{(1)}$
and $\alpha_{ZZ}^{(2)}$ are equal and given by \cite{10}
\begin{align}
   \alpha_{ZZ}^{(1)} = \alpha_{ZZ}^{(2)} = \frac{\alpha_\perp + \alpha_\parallel}{2}
                   +\frac{\alpha_\parallel - \alpha_\perp}{2}\cos 2\theta, \label{alphaZZ1}
\end{align}
%For the orientation with $\xi \parallel Y$, one has
%\begin{align}
%    = \frac{\alpha_\perp + \alpha_\parallel}{2}
%                   +\frac{\alpha_\parallel - \alpha_\perp}{2}\cos 2\theta, \label{alphaZZ2}
%\end{align}
while with $\xi\parallel Z$ one has
\begin{align}
   \alpha_{ZZ}^{(3)} = \alpha_\perp, \label{alphaZZ3}
\end{align}
independent of $\theta$.  In the case of meroaxial disorder, the average polarizability is given by
\begin{align}
   \alpha_\text{ZZ} = \frac{1}{3}\sum_{i=1}^3 \alpha_\text{ZZ}^{(i)} = a + \frac{2}{3}b\cos 2\theta \label{alphaZZ}
\end{align}
where we have defined
\begin{xalignat}{2}
   a = \frac{\alpha_\parallel + 2\alpha_\perp}{3}, & & b  = \frac{\alpha_\parallel - \alpha_\perp}{2}.
\end{xalignat}
Hence the time-dependent correlation function reads 
\begin{align}
   \langle \alpha_{ZZ}(t) \alpha_{ZZ}(0) \rangle =
    a^2 + \frac{4b^2}{9}\bigl\langle \cos 2\theta(t) \cos 2\theta(0)\bigr \rangle. \label{avalphaalphaZZ}
\end{align}
%$N$ is the number of Sc$_\text{2}$C$_\text{2}$ endohedral molecules. The brackets
%$\langle\,\rangle$ stand for a thermal average.

Similarly we obtain for incident radiation in $Z$
direction and scattered radiation in $Y$ direction
\begin{align}
   R_{ZYZY}(\omega) = \frac{N}{2\pi}\int_{-\infty}^{+\infty} dt\,
   e^{i\omega t}\langle\alpha_{ZY}(t)\alpha_{ZY}(0)\rangle. \label{RZYZY}
\end{align}
If the C$_\text{84}$ molecule is in standard orientation,
\begin{align}
   \alpha_{ZY}^{(1)} = b\sin 2\theta, \label{alphaZY1}
\end{align}
while for $\xi \parallel Y$, $\alpha_{ZY}^{(2)} = 0$, and $\xi \parallel Z$, $\alpha_{ZY}^{(3)} = 0$.
The average polarizability for the case of meroaxial disorder reads
\begin{align}
   \alpha_{ZY} = \frac{b}{3}\sin 2\theta, \label{alphaZY}
\end{align}
and the correlation function becomes
\begin{align}
   \langle \alpha_{ZY}(t) \alpha_{ZY}(0) \rangle =
    \frac{b^2}{9}\bigl\langle \sin 2\theta(t) \sin 2\theta(0)\bigr \rangle. \label{avalphaalphaZY}
\end{align}

The problem of determining the scattering laws $R_{ZZZZ}(\omega)$ and $R_{ZYZY}(\omega)$ consists in the
calculation of the orientation-orientation thermal correlation functions
\begin{align}
   C(t) & = \langle\cos 2\theta(t)\cos 2\theta(0)\rangle, \label{Ctdef} \\
   S(t) & = \langle\sin 2\theta(t)\sin 2\theta(0)\rangle, \label{Stdef}
\end{align}
Taking into account the basic relation Eq.\ (\ref{angles}), we expand in terms of $\cos 2\tau$, $\sin 2\tau$, $\cos 2\nu$ and $\sin
2\nu$ thereby obtaining correlation functions of the form
\begin{align}
   C^\text{cccc}(t) & = \langle\cos 2\tau(t)\cos 2\nu(t)\cos 2\tau(0)\cos 2\nu(0)\rangle, \\
%   C^\text{ssss}(t) & = \langle\sin 2\tau(t)\sin 2\nu(t)\sin 2\tau(0)\sin 2\nu(0)\rangle, \\
   S^\text{cscs}(t) & = \langle\cos 2\tau(t)\sin 2\nu(t)\cos 2\tau(0)\sin 2\nu(0)\rangle,
%   S^\text{scsc}(t) & = \langle\sin 2\tau(t)\cos 2\nu(t)\sin 2\tau(0)\cos 2\nu(0)\rangle,
\end{align}
and similarly for $C^\text{ssss}(t)$ and $S^\text{scsc}(t)$.
Observing that $\tau$ refers to quantum dynamics of C$_\text{2}$ and $\nu$ to classical dynamics of
C$_\text{84}$, we decouple the thermal averages:
\begin{xalignat}{2}
   C^\text{cccc} = Q^\text{cc}(t)F^\text{cc}(t), && C^\text{ssss} = Q^\text{ss}(t)F^\text{ss}(t), \\
   S^\text{cscs} = Q^\text{cc}(t)F^\text{ss}(t), && S^\text{scsc} = Q^\text{ss}(t)F^\text{cc}(t).
\end{xalignat}
Here the correlation functions 
\begin{align}
   Q^\text{cc}(t) & = \langle\cos 2\tau(t)\cos 2\tau(0)\rangle, \\
   Q^\text{ss}(t) & = \langle\sin 2\tau(t)\sin 2\tau(0)\rangle,
\end{align}
describe the quantum dynamics of the C$_\text{2}$ unit while the correlation functions
\begin{align}
   F^\text{cc}(t) & = \langle\cos 2\nu(t)\cos 2\nu(0)\rangle, \label{Fcct} \\
   F^\text{ss}(t) & = \langle\sin 2\nu(t)\sin 2\nu(0)\rangle, \label{Fsst}
\end{align}
describe the classical dynamics of the surrounding C$_\text{84}$ molecule.
Finally quantum and classical dynamics occur as products of correlation functions:
\begin{align}
   C(t) = Q^\text{cc}(t)F^\text{cc}(t) + Q^\text{ss}(t)F^\text{ss}(t), \label{Ct} \\
   S(t) = Q^\text{ss}(t)F^\text{cc}(t) + Q^\text{cc}(t)F^\text{ss}(t). \label{St}
\end{align}
Defining Fourier transforms
\begin{align}
   Q(\omega) = \frac{1}{2\pi}\int_{-\infty}^{+\infty}dt\, e^{i\omega t}Q(t), \\
   F(\omega) = \frac{1}{2\pi}\int_{-\infty}^{+\infty}dt\, e^{i\omega t}F(t), \label{Fomega}
\end{align}
and using Eqs.\ (\ref{Ct}), (\ref{St}), (\ref{avalphaalphaZZ}) and (\ref{avalphaalphaZY}), we rewrite the Raman scattering law
in terms of convolutions of Fourier-transformed quantum and classical correlation functions, thereby obtaining
\begin{align}
  R_{ZZZZ}(\omega) = N\bigl[a^2\delta(\omega)
                   +\frac{4b^2}{9}C(\omega)\bigr], \label{RZZZZ2}
\end{align}
with scattering function
\begin{align}
   C(\omega) = \int_{-\infty}^{+\infty}d\omega'\, \bigl[Q^\text{cc}(\omega - \omega')F^\text{cc}(\omega')
                                                       +Q^\text{ss}(\omega - \omega')F^\text{ss}(\omega')\bigr],
						       \label{Comega}
\end{align}
and
\begin{align}
  R_{ZYZY}(\omega) = N\frac{b^2}{9}S(\omega), \label{RZYZY2}
\end{align}
with scattering function
\begin{align}
   S(\omega) = \int_{-\infty}^{+\infty}d\omega'\, \bigl[Q^\text{ss}(\omega - \omega')F^\text{cc}(\omega')
                                                       +Q^\text{cc}(\omega - \omega')F^\text{ss}(\omega')\bigr].
						       \label{Somega}
\end{align}
The first term in brackets on the right-hand side of Eq.\ (\ref{RZZZZ2}) corresponds to the unshifted Rayleigh line of the
spectrum while the function $C(\omega)$ (as also $S(\omega)$ in Eq.\ (\ref{RZYZY2})) accounts for the inelastic part.
Expressions (\ref{Comega}) and (\ref{Somega}) which are convolutions in Fourier space
show that the quantum motion of the C$_\text{2}$ rotor is modulated
by the classical rotational motion of the surrounding C$_\text{84}$ cage.
This is an example of ``direct coupling" of two motions through the detection process \cite{YvinecPick},
in contradistinction to the
``indirect coupling" through a Hamiltonian.  The origin of the direct coupling here is due to the fact that the
detection angle $\theta$ is a sum of two terms, Eq.\ (\ref{angles}).

In the next section we will calculate the quantum mechanical and classical orientational correlation functions for
C$_\text{2}$ and C$_\text{84}$ respectively.

%-----------------------------
\section{Dynamic correlations}\label{secDynamic}
%-----------------------------
\subsection{C$_\text{2}$ quantum rotor}
\begin{figure}
\resizebox{10cm}{!}{\includegraphics{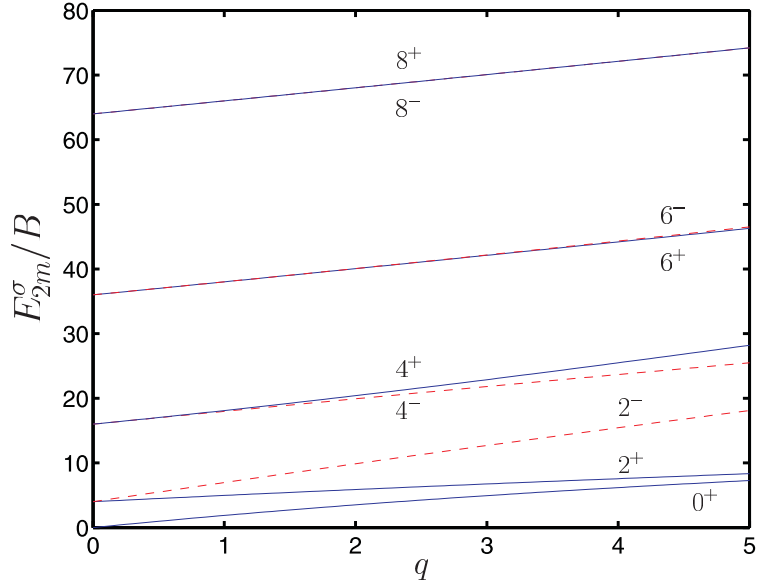}}
\caption{Energy levels of the C$_\text{2}$ planar quantum rotor in the fourfold
molecular potential as a function of the potential strength $q$ (dimensionless
units).  Only levels up to $2m = 8$ are shown.
}
\label{tunneling}
\end{figure}

The quantum mechanics of a diatomic molecular rotor in crystals goes
back to Pauling \cite{7}.  A still valid review of the subject of single particle rotations in molecular
crystals has been given by W. Press \cite{8}.  We will calculate the orientational autocorrelation functions $Q^\text{cc}$ and $Q^\text{ss}$ by starting from the
model of the C$_\text{2}$ planar quantum rotor in the fourfold potential due to the C$_\text{84}$ cage.  We will
refer to this motion as rotational tunneling \cite{8,Hueller}.
We will show that the resonances of the correlation functions $Q^{cc}(\omega)$ and $Q^{ss}(\omega)$ are due to
transitions between tunneling energy levels.
The sole degree
of freedom is the angle $\tau$ which accounts for the rotatory motion of C$_\text{2}$ with respect to the cage.  The
corresponding Schr\"{o}dinger equation reads
\begin{align}
   \left[-B \frac{d^2}{d\tau^2} + \frac{V_0}{2}(1-4\cos 4\tau)\right]\psi(\tau) = E\psi(\tau). \label{Schroedinger}
\end{align}
Here $B = \hbar^2/2I$ is the rotational constant and $I$ the moment of inertia of C$_\text{2}$, $V_0$ is the
barrier height of the potential.
The rotational constant has the dimension of an energy, from experiment \cite{6} one deduces $B = 1.73$ cm$^{-1}$
(wave number units) and $V_0 = 8B$.  These values are supported by ab initio density functional calculations
\cite{6}.
Equation (\ref{Schroedinger}) which is an extension of Mathieu's equation
\cite{7,12} is
also called Hill's equation \cite{13}.  With the definitions
\begin{xalignat}{2}
   \alpha = \frac{1}{B}\left(E-\frac{V_0}{2}\right), && q = \frac{V_0}{4B},
\end{xalignat}
Eq.\ (\ref{Schroedinger}) reads
\begin{align}
   \left[\frac{d^2}{d\tau^2} + \alpha + 2q\cos 4\tau\right]\psi(\tau) = 0. \label{Schroedinger2}
\end{align}
From symmetry considerations (nuclear spin is zero for $^\text{12}$C, electron wave function of
C$_\text{2}^{2-}$ is totally symmetric), it follows that the rotational wave function $\psi(\tau)$ must be
symmetric with respect to the operation $\tau\longrightarrow\tau - \pi$.  For even 
%Inversion symmetry of C$_\text{2}$ implies that the solution must be invariant with respect to the operation
%$\tau\longrightarrow \tau - \pi$.  For even
periodic solutions one makes the ansatz
\begin{align}
   \psi^+(\tau) = \sum_{m=0}^\infty A_{2m}\cos(2m\tau),
\end{align}
$m=0,1,2,\hdots$.  Equation (\ref{Schroedinger2}) then leads to an infinite system of homogeneous equations for
the coefficients $A_{2m}$.  Truncation of this system for a given value $m=N$ leads to $N+1$ equations which
separate into two systems: a first one for $\bigl\{A_0,A_4,\hdots,A_{2N}\bigr\}$ and a second one for
$\bigl\{A_2,A_6,\hdots,A_{2N-2}\bigr\}$ (we take $N$ even).  Solving for the two discriminants yields the roots
$\alpha^+_{2m}(q)$ for $m=0,2,\hdots,N$ and $m=1,3,\hdots,N-1$.  In case of zero potential, i.e.\ $q=0$, these
solutions reduce to the free planar rotor energies $\bigl\{E^+_{2m}(q=0)\bigr\} =
\bigl\{0,\hdots,(2m)^2B,\hdots\bigr\}$ with normalized
eigenfunctions
\begin{align}
   \bigl\{\psi^+_{2m}(\tau)\bigr\} =
   \left\{\frac{1}{\sqrt{\pi}},\hdots,\frac{\cos(2m\tau)}{\sqrt{\pi/2}},\hdots\right\}
\end{align}
in the interval $0\le\tau\le\pi$.  The ansatz for odd periodic solutions reads
\begin{align}
   \psi^-(\tau) = \sum_{m=1}^\infty B_{2m}\sin(2m\tau).
\end{align}
Proceeding as before one determines the roots $\alpha_{2m}^-(q)$.  In case of zero potential the eigenfunctions
are
\begin{align}
   \bigl\{\psi^-_{2m}(\tau)\bigr\} = \left\{\frac{\sin
   2\tau}{\sqrt{\pi}},\hdots,\frac{\sin(2m\tau)}{\sqrt{\pi/2}},\hdots\right\}.
\end{align}
In the following we will label the energy eigenfunctions and eigenvalues by the double index $(\sigma,2m)$,
$\sigma=\pm$,
also in the case of nonzero potential.
In Fig.\ \ref{tunneling} we show plots of $\frac{E_{2m}^\sigma(q)}{B} = \alpha_{2m}^\sigma(q) + 2q$.

%---------
% Figure 
%---------

We next perform a spectral decomposition of the correlation functions $Q^\text{cc}(t)$ and $Q^\text{ss}(t)$
in terms of eigenfunctions and eigenvalues of the Schr\"{o}dinger equation (\ref{Schroedinger}).  In general form
the result reads
\begin{align}
   Q^\text{cc}(t) = \frac{1}{Z}\sum_{i,j}e^{-E_i/T}|C_{ij}|^2e^{i(E_i-E_j)t/\hbar}, \label{Qcct} \\
   Q^\text{ss}(t) = \frac{1}{Z}\sum_{i,j}e^{-E_i/T}|S_{ij}|^2e^{i(E_i-E_j)t/\hbar}, \label{Qsst}
\end{align}
where
\begin{align}
   C_{ij} = \langle i|\cos 2\tau |j\rangle, \\
   S_{ij} = \langle i|\sin 2\tau |j\rangle.
\end{align}
Here the label $i$ ($j$) stands for the double index $(\sigma,2m)$ of the solutions of the
Schr\"{o}dinger
equation.  We calculate the matrix elements $C_{ij}$ and $S_{ij}$ with the free planar rotor
energies.  Symmetry implies that only functions of a same parity $(+,+)$ or $(-,-)$ contribute to C$_{ij}$
while $S_{ij}$ differs from zero only for functions $i,j$ with different parity.
For instance
\begin{align}
   C^{++}_{2m2n} = \int_0^\pi d\tau\, \frac{\cos(2m\tau)}{\sqrt{\pi/2}}\cos 2\tau\frac{\cos(2n\tau)}{\sqrt{\pi/2}}
                 = \frac{1}{2}\delta_{m,n\pm 1}, \label{sr1} \\
%   C^{--}_{2m2n} = \int_0^\pi d\tau\, \frac{\sin(2m\tau)}{\sqrt{\pi/2}}\cos 2\tau\frac{\sin(2n\tau)}{\sqrt{\pi/2}}
%                 = \frac{1}{2}\delta_{m,n\pm 1}. \label{sr2} \\
   S^{+-}_{2m2n} = \int_0^\pi d\tau\, \frac{\cos(2m\tau)}{\sqrt{\pi/2}}\sin 2\tau\frac{\sin(2n\tau)}{\sqrt{\pi/2}}
                 = \frac{1}{2}\delta_{m,n\pm 1}, \label{sr3}
\end{align}
These matrix elements imply selection rules for transitions between energy levels.
%One also has
%%\begin{align}
%$   \bigl\langle\cos 2\tau(t)\sin 2\tau(0)\bigr\rangle = \bigl\langle 2\sin\tau(t)\cos 2\tau(0)\bigr\rangle = 0$.
%%\end{align}
We take Fourier transforms of
Eqs.\ (\ref{Qcct}) and (\ref{Qsst}), using the identity
\begin{align}
   \frac{1}{2\pi}\int_{-\infty}^{+\infty}dt\, e^{i\omega t}e^{i(E_i - E_j)t/\hbar} =
   \delta\left(\omega-\left(\frac{E_j - E_i}{\hbar}\right)\right).
\end{align}
We insert the energies
\begin{align}
   E^\sigma_{2m} = B\alpha^\sigma_{2m}(q) + \frac{V_0}{2},
\end{align}
and take into account the selection rules (\ref{sr1}) -- (\ref{sr3}).  Defining the frequency transfer
\begin{align}
   \omega^{\sigma\sigma'}_{mn} = \frac{E_{2n}^{\sigma'} - E_{2m}^{\sigma}}{\hbar}, 
\end{align}
we obtain
\begin{multline}
   Q^\text{cc}(\omega) = \frac{1}{2Z}\Biggl\{e^{-E^+_0/T}\delta(\omega - \omega^{++}_{01})
                                     +\sum_{m=1}^\infty\frac{e^{-E^+_{2m}/T}}{2}\bigl[\delta(\omega -
				     \omega^{++}_{mm+1}) + \delta(\omega - \omega^{++}_{mm-1})\bigr] \\
				     +\frac{e^{-E^-_{2}/T}}{2}\bigl[\delta(\omega - \omega^{--}_{12})
				     +\sum_{m=2}^\infty\frac{e^{-E^-_{2m}/T}}{2}\bigl[\delta(\omega -
				     \omega^{--}_{mm+1}) + \delta(\omega - \omega^{--}_{mm-1})\bigr]\Biggr\}
				     \label{Qccomega},
\end{multline}
with
\begin{align}
   Z = e^{-E^+_0/T} + \sum_{m=1}^\infty \left(e^{-E^+_{2m}/T} + e^{-E^-_{2m}/T}\right). \label{partitionsum}
\end{align}
Similarly we get
\begin{multline}
   Q^\text{ss}(\omega) = \frac{1}{2Z}\Biggl\{e^{-E^+_0/T}\delta(\omega - \omega^{+-}_{01}) \\
                                     +\frac{e^{-E^+_{2}/T}}{2}\delta(\omega - \omega^{+-}_{12})
				     +\sum_{m=2}^\infty\frac{e^{-E^+_{2m}/T}}{2}\bigl[\delta(\omega -
				     \omega^{+-}_{mm+1}) + \delta(\omega - \omega^{+-}_{mm-1})\bigr] \\
				     +\sum_{m=1}^\infty\frac{e^{-E^-_{2m}/T}}{2}\bigl[\delta(\omega -
				     \omega^{-+}_{mm+1}) + \delta(\omega - \omega^{-+}_{mm-1})\bigr]\Biggr\}.
				     \label{Qssomega}
\end{multline}
We notice that in absence of the uniaxial rotation of the C$_\text{84}$ cage, i.e.\ for $\nu = 0$, the
correlation functions Eqs.\ (\ref{Fcct}) and (\ref{Fsst}) reduce to constants: $F^\text{cc} = 1$,
$F^\text{ss} = 0$.  Hence the spectral functions $C(\omega)$ and $S(\omega)$ entering the Raman scattering laws
Eqs.\ (\ref{RZZZZ2}) and (\ref{RZYZY2}) reduce to $Q^\text{cc}(\omega)$ and $Q^\text{ss}(\omega)$ and exhibit
infinitely sharp $\delta$-peaks which account for transitions between quantized planar rotor states.
In Table I we have quoted some values ($m\le 4$) of $\omega_{m n}^{\sigma \sigma'}$ for $q = 0$ (free rotor) and $q = 2$
(value of the potential strength taken from experiment in Ref.\ \onlinecite{6}).

%--------
% Table 1
%--------
\begin{table}
\caption{Tunneling frequency transfers
$\omega_{m n}^{\sigma \sigma'}$, $n = m\pm 1$, $\sigma,\sigma'=\pm$, in units cm$^{-1}$.
}
\label{hbaromega}
\begin{ruledtabular}
\begin{tabular}{rrrrrrrrr}
$m$ & $\omega_{mm+1}^{++}$ & $\omega_{mm+1}^{--}$ & $\omega_{mm+1}^{+-}$ & $\omega_{mm+1}^{-+}$
    & $\omega_{mm-1}^{++}$ & $\omega_{mm-1}^{--}$ & $\omega_{mm-1}^{+-}$ & $\omega_{mm-1}^{-+}$ \\
\hline
\multicolumn{9}{c}{$q=0$} \\
\hline
$0$ &  $6.92$ &          & $6.92$ &          &         &         &         &         \\
$1$ & $20.76$ & $20.76$ & $20.76$ & $20.76$ &     $-6.92$ &         &         & $-6.92$ \\
$2$ & $34.60$ & $34.60$ & $34.60$ & $34.60$ &    $-20.76$ & $-20.76$ & $-20.76$ & $-20.76$ \\
$3$ & $48.44$ & $48.44$ & $48.44$ & $48.44$ &    $-34.60$ & $-34.60$ & $-34.60$ & $-34.60$ \\
$4$ & $62.28$ & $62.28$ & $62.28$ & $62.28$ &    $-48.44$ & $-48.44$ & $-48.44$ & $-48.44$ \\
\hline
\multicolumn{9}{c}{$q=2$} \\
\hline
$0$ &  $4.10$ &          & $10.99$ &          &         &         &         &         \\
$1$ & $25.12$ & $17.39$ & $24.28$ & $18.23$ &     $-4.10$ &         &         & $-10.99$ \\
$2$ & $34.00$ & $34.87$ & $34.02$ & $34.84$ &    $-25.12$ & $-17.39$ & $-18.23$ & $-24.28$ \\
$3$ & $48.40$ & $48.38$ & $48.40$ & $48.38$ &    $-34.00$ & $-34.87$ & $-34.84$ & $-34.02$ \\
$4$ & $62.26$ & $62.26$ & $62.26$ & $62.26$ &    $-48.40$ & $-48.38$ & $-48.38$ & $-48.40$ \\
%\hline
%\multicolumn{9}{c}{$q=2$} \\
%\hline
%$0$ &  $-5.90$ &          & $-15.82$ &          &         &         &         &         \\
%$1$ & $-36.14$ & $-25.01$ & $-34.93$ & $-26.23$ &  $5.90$ &         &         & $15.85$ \\
%$2$ & $-48.91$ & $-50.16$ & $-48.95$ & $-50.16$ & $36.14$ & $25.01$ & $26.23$ & $25.01$ \\
%$3$ & $-69.64$ & $-69.60$ & $-69.64$ & $-69.60$ & $48.91$ & $50.16$ & $50.13$ & $50.16$ \\
%$4$ & $-89.58$ & $-89.58$ & $-89.58$ & $-89.58$ & $69.64$ & $69.60$ & $69.60$ & $69.60$ \\
\end{tabular}
\end{ruledtabular}
\end{table}

\subsection{C$_\text{84}$ uniaxial rotational diffusion}\label{s3B}
In order to calculate the classical correlation functions $F^{cc}(t)$ and $F^{ss}(t)$ we
treat the C$_\text{84}$ molecule as a classical uniaxially diffusing rotor with rotation axis $S_4$ in coincidence with a
cubic crystal axis, in casu the $X'$ axis.  The corresponding rotation angle $\nu$ is measured away from the $Z'$
axis.
Equivalently one considers the $S_4$ axis along $Y'$ and $Z'$ (meroaxial disorder).
Given the $S_4$ axis it would be tempting to study this
motion in a crystal field potential of fourfold symmetry.  Such a study can be carried out along the lines of
Ref.\ \onlinecite{14} and leads to a continued fraction expansion in terms of frequency moments of the orientational
variables.  It is adequate in the case of a strong crystal field potential since then one can limit the continued
fraction to a few steps.  However this approximation is not valid in the case of weak potentials.  Since
the equator of the C$_\text{84}$ molecule for rotations about $S_4$ deviates only slightly from circular
shape, we prefer to consider the rotator about the $S_4$ axis in the rotational-diffusion
approximation.  This model has the obvious advantage of simplicity and leads to a linear temperature-dependent
broadening of the tunneling transition lines.  Within the uniaxial diffusion model
the C$_\text{84}$ molecule experiences a random rotational torque (also called Brownian motion
torque) about its $S_4$ axis.  This torque is caused by the thermal motion of the surrounding lattice (heat bath).
In that respect the present problem is different from the situation of the heavy symmetrical top with
gravitational torque since on the molecular scale the effect of gravitation is negligible in comparison with the
heat bath.

  The idea of rotational diffusion goes back to Debye \cite{15} who applied the concept of rotational
Brownian motion to the theory of dielectric relaxation (see also \cite{8} and \cite{16}).  In Appendix \ref{appA} we give some details for the present
problem.
As a result we obtain 
\begin{align}
   F^\text{cc}(t) = \bigl\langle\cos 2\nu(t) \cos 2\nu(0)\bigr\rangle = \frac{1}{2}e^{-4D_\text{R}t},
   \label{Fcctb} \\
   F^\text{ss}(t) = \bigl\langle\sin 2\nu(t) \sin 2\nu(0)\bigr\rangle = \frac{1}{2}e^{-4D_\text{R}t}. \label{Fsstb}
\end{align}
Here the rotational diffusion coefficient $D_\text{R}$ is given by the Einstein
relation
\begin{align}
   D_\text{R} = \frac{k_\text{B}T}{\zeta}, \label{diffcoeff}
\end{align}
where $\zeta$ is the friction coefficient and $T$ the temperature.
The equality of $F^\text{cc}(t)$ and $F^\text{ss}(t)$ is a consequence of our neglect of the crystal field
potential within the large-friction approximation.  From Eqs.\ (\ref{Ct}) and (\ref{St}) one sees that then 
\begin{align}
   C(t) = S(t).  \label{alsosees}
\end{align}
In the following we will neglect the superscripts $ss$ and $cc$ on $F^{cc}$ and $F^{ss}$ and write just $F$.

\begin{figure}
\subfigure{\resizebox{8cm}{!}
{\includegraphics{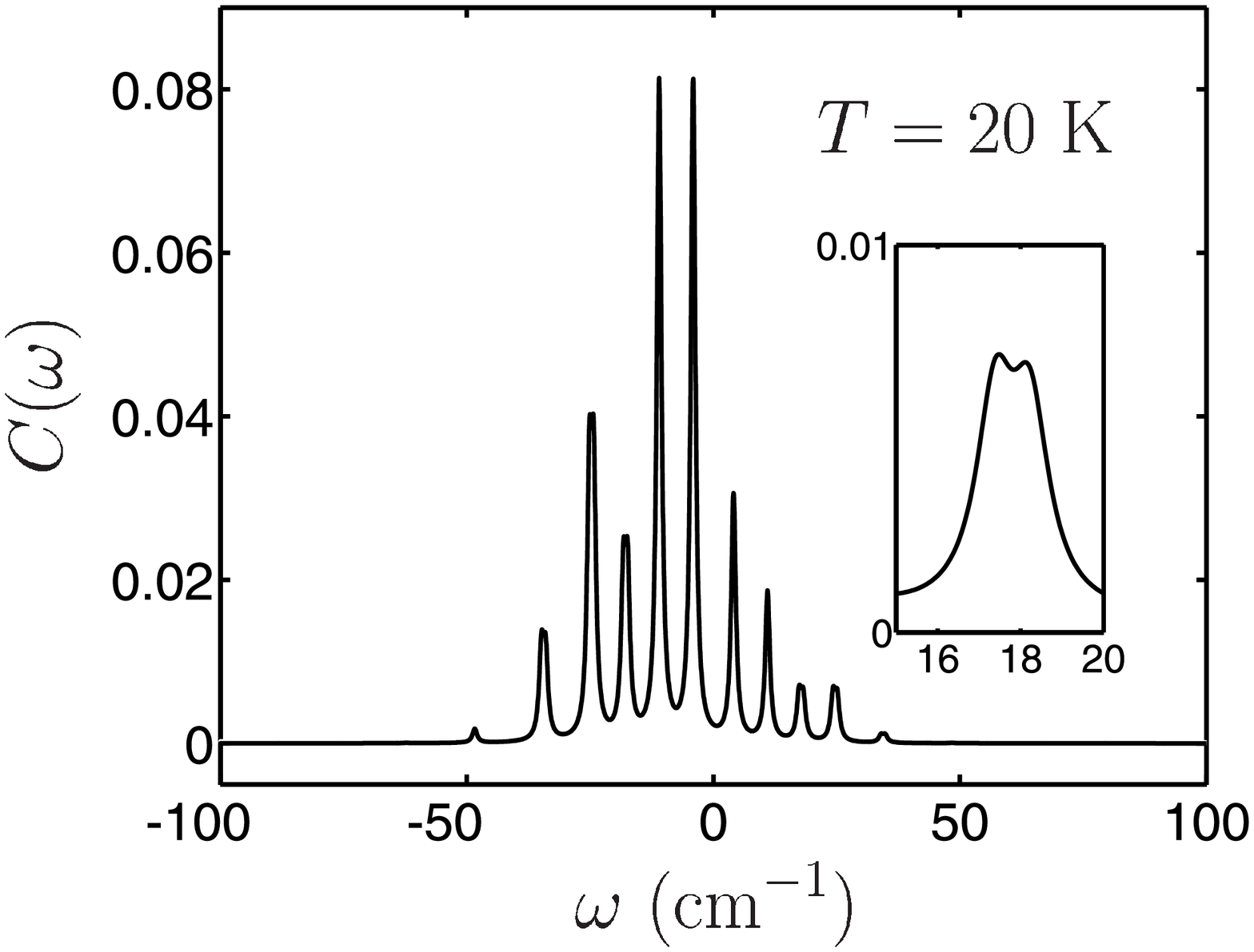}}} \\
\subfigure{\resizebox{8cm}{!}
{\includegraphics{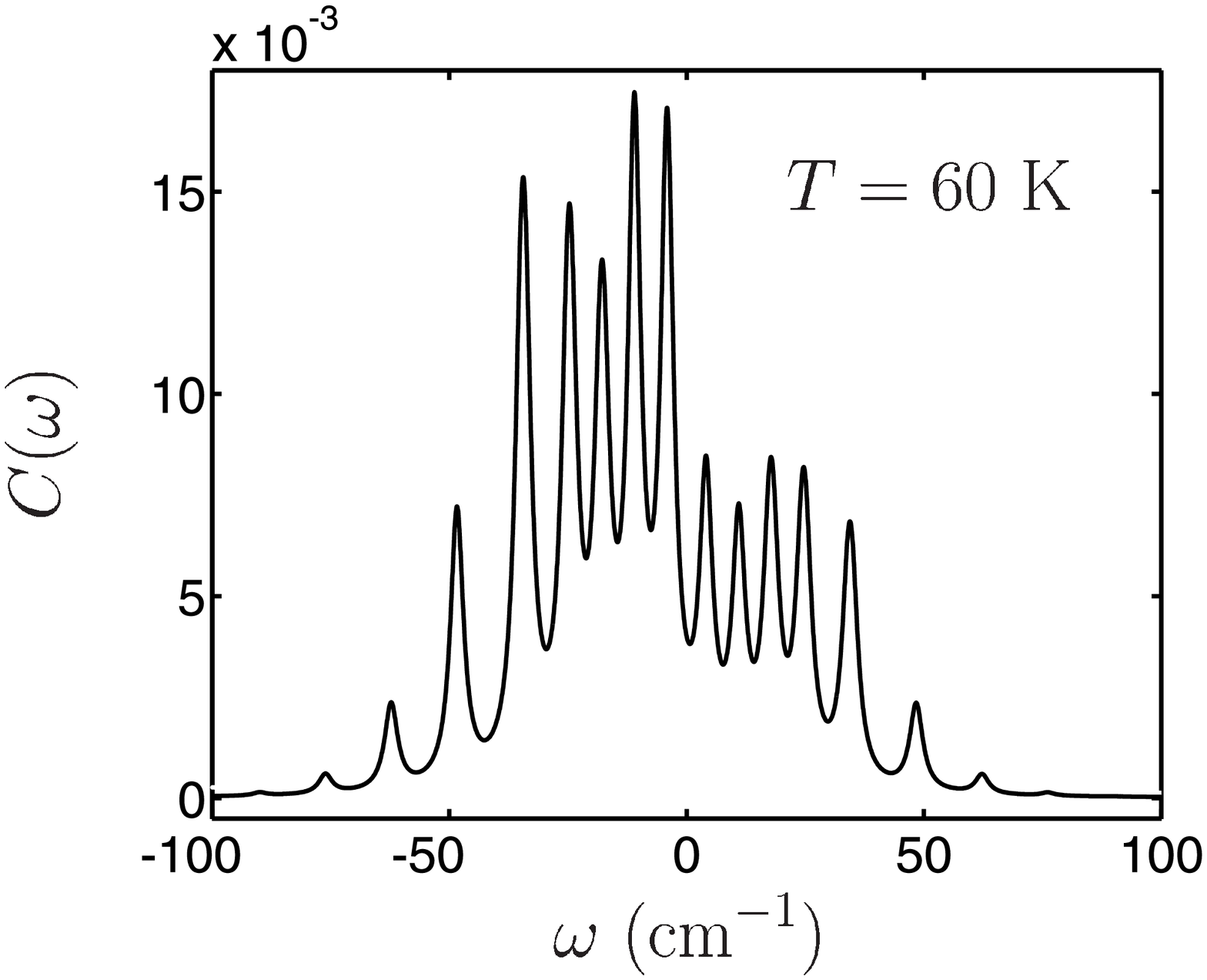}}} \\
\subfigure{\resizebox{8cm}{!}
{\includegraphics{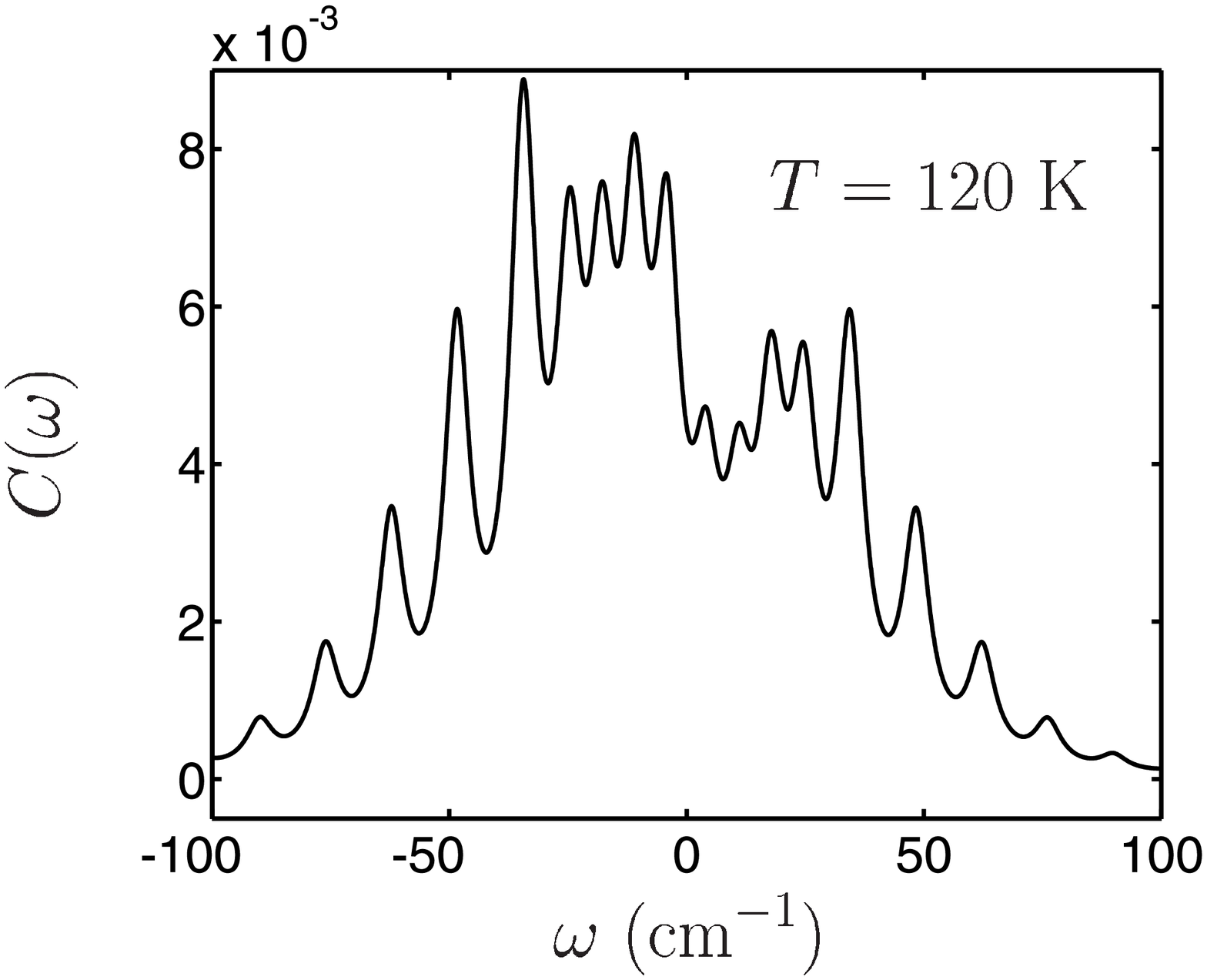}}}
\caption{Scattering
function $C(\omega)$ for $T=20$ K, $60$ K and $120$ K.  The width $8D_\text{R}$ is the same for all resonance
lines and increases linearly
from $1.11$ cm$^{-1}$ to $6.66$ cm$^{-1}$ for $T=20$ K and $120$ K, respectively.
The inset at $T=20$ K shows the splitting of the $\omega_{12}^{--}$ and $\omega_{12}^{+-}$ resonances.
}
\label{Comegaplot}
\end{figure}

The Fourier transform is obtained from Eq.\ (\ref{Fsstb}) with the result
\begin{align}
   F(\omega) = \frac{1}{2\pi}\left[\frac{4D_R}{\omega^2 + 16D_\text{R}^2}\right].
   \label{resFccomega}
\end{align}
We rewrite the scattering functions Eqs.\ (\ref{Comega}) and (\ref{Somega}) as
\begin{align}
   C(\omega) = S(\omega) = \int_{-\infty}^{+\infty}d\omega'\,\bigl[Q^\text{ss}(\omega-\omega') +
   Q^\text{cc}(\omega-\omega')\bigr]F(\omega'). \label{integral}
\end{align}
%We find that the modulation of the tunneling motion of the Sc$_\text{2}$C$_\text{2}$ rotor by the classical
%diffusion rotation of the C$_\text{84}$ cage about the $S_4$ axis leads to equality of the spectral functions $S(\omega)$
%and $C(\omega)$.
Using expressions (\ref{Qccomega}) and (\ref{Qssomega}) we carry out the integral over $\omega'$ and obtain
\begin{align}
   C(\omega) = C_{++}(\omega) + C_{--}(\omega) + C_{+-}(\omega) + C_{-+}(\omega) \label{resComega}
\end{align}
where
\begin{align}
   C_{++}(\omega) & =  \frac{1}{2Z}\left\{e^{-E_0^+/T}F(\omega - \omega_{01}^{++})
    + \sum_{m=1}^\infty \frac{e^{-E_{2m}^+/T}}{2}\bigl[F(\omega - \omega_{mm+1}^{++})
     + F(\omega - \omega_{mm-1}^{++})\bigr]\right\}, \label{Cpp} \\
   C_{--}(\omega) & =  \frac{1}{2Z}\left\{\frac{e^{-E_2^-/T}}{2}F(\omega - \omega_{12}^{--})
    + \sum_{m=2}^\infty \frac{e^{-E_{2m}^-/T}}{2}\bigl[F(\omega - \omega_{mm+1}^{--})
     + F(\omega - \omega_{mm-1}^{--})\bigr]\right\}, \\
   C_{+-}(\omega) & =  \frac{1}{2Z}\left\{e^{-E_0^+/T}F(\omega - \omega_{01}^{+-})
    + \frac{e^{-E_2^+/T}}{2}F(\omega - \omega_{12}^{+-})\right. \nonumber \\
    & \phantom{=\frac{1}{2Z}\left\{\right.}\left. + \sum_{m=2}^\infty \frac{e^{-E_{2m}^+/T}}{2}\bigl[F(\omega - \omega_{mm+1}^{+-})
     + F(\omega - \omega_{mm-1}^{+-})\bigr]\right\}, \\
   C_{-+}(\omega) & =  \frac{1}{2Z}\left\{
    \sum_{m=1}^\infty \frac{e^{-E_{2m}^-/T}}{2}\bigl[F(\omega - \omega_{mm+1}^{-+})
     + F(\omega - \omega_{mm-1}^{-+})\bigr]\right\}. \label{Cmp}
\end{align}
%From Eq.\ (\ref{resFccomega}) for $F(\omega)$
We see that $C(\omega)$ is a sum of weighted Lorentzians
\begin{align}
   F(\omega - \omega_{mm\pm 1}^{\sigma\sigma'}) = \frac{1}{2\pi}\left[ \frac{4D_\text{R}}{(\omega - \omega_{mm\pm
   1}^{\sigma\sigma'})^2 + 16D_\text{R}^2} \right] \label{Lorentzian}
\end{align}
centered
around the allowed frequency transfers $\omega=\omega^{\sigma\sigma'}_{mm\pm 1}$ and of width
$8D_\text{R}$ (full width half maximum).
Since $D_\text{R}$ has dimension s$^{-1}$, it follows from Eq.\ (\ref{diffcoeff})
that $\zeta$ has the dimension of an action. 
We write $\zeta = \zeta_n h$, where $\zeta_n$ is a dimensionless number taken as parameter.  We then obtain
$D_\text{R} = 2.08\times 10^{10}(T/\zeta_n)$ s$^{-1}$ where $T$ is measured in Kelvin.  Equivalently,
$D_\text{R}=0.694(T/\zeta_n)$ cm$^{-1}$.
Since to our knowledge there are so far no direct measurements of the orientational dynamics of the C$_\text{84}$
molecule in Sc$_\text{2}$C$_\text{2}$@C$_\text{84}$, we will choose a value of $D_\text{R}$ such that the
correlation time $\tau_c = (4D_\text{R})^{-1}$ has a value that is intermediate between the values of $2$ ns and
$5$ ps measured by NMR experiments for the C$_\text{70}$ molecule in the low-temperature monoclinic and
high-temperature fcc phases of solid C$_\text{70}$, respectively \cite{Tycko}.
Assuming that $\zeta_n=100$ is a realistic value (then $D_\text{R} =
10^{10}$ s$^{-1}$ at $T = 50$ K), we have plotted the scattering function $C(\omega)$ for several temperatures in
Fig.\ \ref{Comegaplot}.  The resonances are centered at the frequency transfers $\omega_{mn}^{\sigma\sigma'}$ for
the potential strength $q=2$.  The spectra reflect the characteristic assymetries for $\omega>0$ and $\omega<0$
due to anti-Stokes and Stokes processes, respectively.  In our calculations, we have included transitions with
the values $m,n=0,\hdots,19$.
% in Table \ref{hbaromega}.

%For numerical estimates we assume that the friction coefficient $\zeta$ in Eq.\ (\ref{diffcoeff}) has a value
%$\zeta=200\hbar$.  We then get $D_\text{R}=6.57\times 10^8\ T$ s$^{-1}$ which corresponds to $2.19\times 10^{-2}\ T$ cm$^{-1}$
%($T$ in units K).
%In Fig.\ \ref{Comegaplot} we have plotted the function $C(\omega)$ for several temperatures.

%---------
% Figure 5
%---------

We notice that if one artificially excludes the tunneling motion of the C$_\text{2}$ unit by taking a fixed
value, say $0$, for the angle $\tau$, one finds that $Q^{ss}(\omega)=0$ and $Q^{cc} = \delta(\omega)$.  Then Eq.\
(\ref{integral}) becomes $C(\omega) = F(\omega)$.  Since the C$_\text{2}$-unit is dragged along with the classical
rotational diffusion of the encapsulating C$_\text{84}$ molecule, its polarizability is changing accordingly with
time and
the Raman scattering laws $R_{ZZZZ}(\omega)$ and $R_{ZYZY}(\omega)$ will exhibit a
Lorentzian $F(\omega)$ of width $4D_\text{R}$ centered at $\omega = 0$.

%------------------------
\section{Powder averages}\label{secPowder}
%------------------------
In Sect.\ \ref{secDynamic} we have considered
a cubic crystal with crystal axes $(X',Y',Z')$ in coincidence with the laboratory-fixed cubic axes $(X,Y,Z)$.  
Since experiments are performed on powder samples, we will extend the previous results.  The powder sample
consists of a large number of arbitrarily oriented cubic crystallites, each crystallite has symmetry $Fm\overline{3}m$ where the
Sc$_\text{2}$C$_\text{2}$@C$_\text{84}$ units are meroaxially disordered \cite{4}.
We first will consider one single crystallite where the crystal-fixed system of axes is related to the
laboratory-fixed system of axes by the Euler angles $(\alpha,\beta,\gamma)$.  The C$_\text{2}$ rotors are now
moving in planes normal to the $X'$, $Y'$, $Z'$ axes of the rotated coordinate system.  This means that the
polarizabilities $\alpha_{ZZ}^R$ or $\alpha_{ZY}^R$ measured in the laboratory-fixed coordinate system will depend on
the Euler angles of the given crystallite.

In Appendix \ref{appB} we have calculated the polarizability components $\alpha_{ZZ}^{(i)R}$ which are
obtained from $\alpha_{ZZ}^{(i)}$ by application of the rotation operation $R(\alpha,\beta,\gamma)$:
\begin{align}
   \alpha_{ZZ}^{(i)R} = R(\alpha,\beta,\gamma)\alpha_{ZZ}^{(i)}.
\end{align}
The meroaxial average
\begin{align}
   \alpha_{ZZ}^R = \frac{1}{3}\sum_{i=1}^3 \alpha_{ZZ}^{(i)R}
\end{align}
is obtained as
\begin{align}
   \alpha_{ZZ}^R = a +
   \frac{b}{3}\sum_{i=1}^3\bigl[A_{ZZ}^{(i)}(\beta,\gamma) + B_{ZZ}^{(i)}(\beta,\gamma)\sin 2\theta +
   C_{ZZ}^{(i)}(\beta,\gamma)\cos 2\theta\bigr]. \label{alphaZZRmeroaxial}
\end{align}
where $i=1$ refers to $\xi\parallel X'$, $i=2$ to $\xi \parallel Y'$ and $i=3$ to $\xi \parallel Z'$.
The coefficients $A_{ZZ}^{(i)}(\beta,\gamma)$, $B_{ZZ}^{(i)}(\beta,\gamma)$ and
$C_{ZZ}^{(i)}(\beta,\gamma)$ are derived in Appendix \ref{appB}, they are found to depend on only two Euler
angles.

In the present section we assume that the meroaxial disorder is static or equivalently there are no
reorientations of the C$_\text{84}$ molecules among the three meroaxial directions in a given crystallite.  The
angle $\theta$ is then the sole dynamic quantity.  The time-dependent polarizability correlation function per
molecule in the given crystallite is obtained as
\begin{align}
   \bigl\langle \alpha_{ZZ}^R(t) \alpha_{ZZ}^R(0)\bigr\rangle
   = a^2 + \frac{b^2}{9}\sum_{i,j}\bigl[ A_{ZZ}^{(i)}A_{ZZ}^{(j)}
     + B_{ZZ}^{(i)}B_{ZZ}^{(j)}S(t)
     + C_{ZZ}^{(i)}C_{ZZ}^{(j)}C(t)
     \bigr]. \label{alphaZZRmeroaxialt}
\end{align}   
The correlation functions $C(t)$ and $S(t)$, defined by Eqs.\ (\ref{Ctdef}) and (\ref{Stdef}) respectively, have been calculated in
Sects.\ II and III.

The powder average for a function $F(\beta,\gamma)$ is defined as
\begin{align}
   \overline{F} =
   \frac{1}{4\pi}\int_0^{2\pi}d\gamma\,\int_0^{\pi}d\beta\,\sin\beta F(\beta,\gamma). \label{Fbetagammaaverage}
\end{align}
The results for the products $\overline{A_{ZZ}^{(i)}A_{ZZ}^{(j)}},
\overline{B_{ZZ}^{(i)}B_{ZZ}^{(j)}},\hdots$ are quoted in Appendix \ref{appB}.  The
powder-averaged polarizability correlation function per molecule reads
\begin{align}
   \overline{\bigl\langle\alpha_{ZZ}^R(t)\alpha_{ZZ}^R(0)\bigr\rangle} & = a^2 +
   \frac{b^2}{9}\left[\frac{8}{15}S(t) + \frac{12}{15}C(t)\right]. \label{refaZZR}
\end{align}
Taking into account $S(t) = C(t)$, Eq.\ (\ref{alsosees}), we obtain the Raman scattering law for a
powder-averaged sample with meroaxial disorder:
\begin{align}
   \overline{R_{ZZZZ}(\omega)} & = N\left(a^2\delta(\omega) + \frac{4}{27}b^2C(\omega)\right).
\end{align}
The expression for a single crystal with meroaxial disorder has been given by Eq.\ (\ref{RZZZZ2}).

In an analogous way one calculates
\begin{align}
   \alpha_{ZY}^R = \frac{1}{3}\sum_{i=1}^3\alpha_{ZY}^{(i)R}
\end{align}
with the result
\begin{align}
   \alpha_{ZY}^R = \frac{b}{3}\sum_{i=1}^3\bigl[A_{ZY}^{(i)}(\alpha,\beta,\gamma)
    + B_{ZY}^{(i)}(\alpha,\beta,\gamma)\sin 2\theta + C_{ZY}^{(i)}(\alpha,\beta,\gamma)
   \cos 2\theta\bigr]. \label{alphaZYRmeroaxial}
\end{align}
The coefficients $A_{ZY}^{(i)},\hdots,C_{ZY}^{(i)}$ are given in Appendix \ref{appB}.
The time-dependent polarizability correlation function per molecule reads
\begin{align}
   \bigl\langle \alpha_{ZY}^R(t)\alpha_{ZY}^R(0)\bigr\rangle
   = \frac{b^2}{9}\sum_{i,j}\bigl[A_{ZY}^{(i)}A_{ZY}^{(j)}
     + B_{ZY}^{(i)}B_{ZY}^{(j)}S(t)
     + C_{ZY}^{(i)}C_{ZY}^{(j)}C(t)
   \bigr].  \label{alphaZYRmeroaxialt}
\end{align}

%The coefficients $A_{ZY}^{(i)},\hdots,C_{ZY}^{(i)}$ given in Appendix \ref{appB} are functions of the three
%Euler angles $\alpha$, $\beta$, $\gamma$.
The powder average of a function $F(\alpha,\beta,\gamma)$ is defined by 
\begin{align}
   \overline{F} =
   \frac{1}{8\pi^2}\int_0^{2\pi}d\alpha\,\int_0^{2\pi}d\gamma\,\int_0^{\pi}d\beta\,\sin\beta F(\alpha,\beta,\gamma).
   \label{Falphabetagammaaverage}
\end{align}

%Assuming again static meroaxial disorder and
Taking into account the powder averages
$\overline{A_{ZY}^{(i)}A_{ZY}^{(j)}}$ etc., calculated in Appendix \ref{appB}, we obtain
\begin{align}
   \overline{\bigl\langle\alpha_{ZY}^R(t)\alpha_{ZY}^R(0)\bigr\rangle} & = \frac{b^2}{9}\left[\frac{11}{15}S(t)
   + \frac{4}{15}C(t)\right]. \label{refaZYR}
\end{align}
The Raman scattering law then reads
\begin{align}
   \overline{R_{ZYZY}(\omega)} & = N\frac{b^2}{9}C(\omega),
\end{align}
where again we have used $S(t) = C(t)$, Eq.\ (\ref{alsosees}).  We see that the powder-averaged
expression is the same as the one for a single crystal
with meroaxial disorder, Eq.\ (\ref{RZYZY2}).

%-----------------------------------
\section{Dynamic Meroaxial Disorder}
%-----------------------------------
So far we have assumed that the orientation of the long axis ($S_4$) of the C$_\text{84}$ molecule in a given
cubic crystallite along the equivalent $\langle 100\rangle$ directions is random but static.
The sole effect of the heat bath was the uniaxial rotational diffusion studied in Sect.\ \ref{s3B}.
While this
situation
of static meroaxial disorder is realistic at temperatures inferior to say $100$ K, it becomes less valid at higher
$T$.
Here again we refer to the situation in solid C$_\text{70}$ where with increasing temperature it is found that
the uniaxial rotation axis flips between different symmetry equivalent orientations such that the rotational
motion becomes more and more isotropic \cite{Dennis,Mcrae,Maniwa,Blinc,Tycko,Christides}.
We therefore will extend the previous model and take into account the situation where a molecule at a given
lattice site in one crystallite changes orientation with the $S_4$ axis jumping randomly between equivalent potential minima in
$\langle 100 \rangle$ directions.  Here the heat bath causes stochastic torques about axes perpendicular to the
long axis of the C$_\text{84}$ molecule or equivalently perpendicular to the rotation axis of the encapsulated
Sc$_\text{2}$C$_\text{2}$ quantum gyroscope.  We recall that accordingly the normal to the plane of the
C$_\text{2}$
quantum rotor will change its orientation.  Within a simple three sites stochastic jump model (see e.g.\ \cite{8}), the conditional
probability $p(i,t|j,0)$ to find a C$_\text{84}$ molecule in an orientation $i=1,2,3$ at time $t\ge 0$ when it was in orientation $j=1,2,3$
at time $0$ is obtained by solving a system of three linear differential equations.  One obtains
\begin{align}
   p(i,t|j,0) & = \frac{1}{3}(1 + 2e^{-3w t})\text{, }i=j, \label{ieqj} \\
   p(i,t|j,0) & = \frac{1}{3}(1 - e^{-3w t}) \text{, }i\ne j, \label{ineqj}
\end{align}
where $w$ is the transition rate for a molecular reorientation.  We associate the
transition rate with the inverse of a relaxation time:
\begin{align}
   w = \frac{1}{\tau} = \frac{1}{\tau_0}e^{-E_\text{a}/T}.
\end{align}
Here we have assumed an Arrhenius-type law, known from reaction rate theory \cite{18,16}, where $1/\tau_0$ is an attempt frequency and $E_\text{a}$ an activation
energy for meroaxial reorientations of the Sc$_\text{2}$C$_\text{2}$@C$_\text{84}$ complex as a whole.
The equilbrium value of the conditional probability is independent of the initial and final orientation and
corresponds to an a priori probability:
\begin{align}
   \lim_{t\longrightarrow \infty} p(i,t|j,0) = \frac{1}{3}.
\end{align}
In the previous section the meroaxial orientations within a given crystallite
have been characterized by the coefficients $\bigl\{
A_{ZZ}^{(i)},B_{ZZ}^{(i)},C_{ZZ}^{(i)}\bigr\}$, $\bigl\{A_{ZY}^{(i)},B_{ZY}^{(i)},C_{ZY}^{(i)}\bigr\}$ in Eqs.\
(\ref{alphaZZRmeroaxial}) and (\ref{alphaZYRmeroaxial}) of the polarizabilities $\alpha_{ZZ}^R$ and
$\alpha_{ZY}^R$.  Treating these coefficients as dynamic stochastic quantities we obtain instead of Eqs.\
(\ref{alphaZZRmeroaxialt}) and (\ref{alphaZYRmeroaxialt})
\begin{align}
   \bigl\langle\alpha_{ZZ}^R(t)\alpha_{ZZ}^R(0)\bigr\rangle & = a^2 + b^2\Bigl[  
   \bigl\langle A_{ZZ}(t)A_{ZZ}(0)\bigr\rangle \nonumber \\
   &\phantom{a^2 + b^2\Bigl[} + \bigl\langle B_{ZZ}(t) B_{ZZ}(0)\bigr\rangle S(t)
   + \bigl\langle C_{ZZ}(t) C_{ZZ}(0)\bigr\rangle C(t)
   \Bigr], \label{aZZ} \\
   \bigl\langle\alpha_{ZY}^R(t)\alpha_{ZY}^R(0)\bigr\rangle & = b^2\Bigl[  
   \bigl\langle A_{ZY}(t)A_{ZY}(0)\bigr\rangle \nonumber \\
   &\phantom{b^2\Bigl[} + \bigl\langle B_{ZY}(t) B_{ZY}(0)\bigr\rangle S(t)
   + \bigl\langle C_{ZY}(t) C_{ZY}(0)\bigr\rangle C(t)
   \Bigr]. \label{aZY}
\end{align}
The correlation functions $\bigl\langle A_{ZZ}(t)A_{ZZ}(0)\bigr\rangle,\hdots,
\bigl\langle C_{ZY}(t)C_{ZY}(0)\bigr\rangle$ which refer to meroaxial reorientations are calculated within the
frame of the stochastic jump model.  For instance for a given set $\bigl\{A^{(i)},i=1,2,3\bigr\}$ one has
\begin{align}
   \bigl\langle A(t)A(0)\bigr\rangle = \frac{1}{3}\sum_{i,j}A^{(i)}A^{(j)}p(i,t|j,0),
\end{align}
where the conditional probabilities $p(i,t|j,0)$ are given by Eqs.\ (\ref{ieqj}) and (\ref{ineqj}), while the factor $1/3$
accounts for the equilibrium initial probability.  Since
the coefficients $A^{(i)}$
depend on the Euler angles which specify the orientation of a given crystallite (Sect.\ \ref{secPowder}), the
powder-averaged correlation functions are obtained by averaging over the Euler angles:
\begin{align}
   \overline{\bigl\langle A(t)A(0)\bigr\rangle} = \frac{1}{3}\sum_{i,j}\overline{A^{(i)}A^{(j)}}p(i,t|j,0).
   \label{AtA0}
\end{align}
Taking into account the values of the powder-averaged products given in Appendix \ref{appB}, we obtain:
\begin{align}
   \overline{\bigl\langle A_{ZZ}(t)A_{ZZ}(0)\bigr\rangle} & = \frac{4}{45}e^{-3t/\tau}, \\
   \overline{\bigl\langle B_{ZZ}(t)B_{ZZ}(0)\bigr\rangle} & = \frac{8}{135}\left[1 + 2e^{-3t/\tau}\right], \\
   \overline{\bigl\langle C_{ZZ}(t)C_{ZZ}(0)\bigr\rangle} & = \frac{12}{135}\left[1 + 3e^{-3t/\tau}\right].
\end{align}
The powder average of Eq.\ (\ref{aZZ}) then reads
\begin{align}
   \overline{\bigl\langle \alpha_{ZZ}^R(t)\alpha_{ZZ}^R(0)\bigr\rangle} = a^2 + \frac{4b^2}{27}D(t),
\end{align}
where the function $D(t)$ is given by
\begin{align}
   D(t) = \bigl[C(t) + \frac{3}{5}e^{-3t/\tau} + \frac{13}{5}C(t)e^{-3t/\tau}\bigr]. \label{Dt}
\end{align}
Here we have used again $C(t) = S(t)$, Eq.\ (\ref{alsosees}).
The first term on the right-hand side $C(t)$ accounts for the superposition of the quantum motion (tunneling) of
the C$_\text{2}$ rotor
and the uniaxial rotational diffusion of the encapsulating C$_\text{84}$ molecule, the second term $\propto e^{-3t/\tau}$ accounts
for the classical motion of the C$_\text{2}$ rotor when its plane of motion is changing with the meroaxial
reorientations of the encapsulating C$_\text{84}$ molecule, finally the third term $\propto C(t)e^{-3t/\tau}$
accounts for the interference of the two motions of the C$_\text{84}$ molecule with the tunneling of the
C$_\text{2}$ rotor.
% In the limit of small relaxation such
%that $\tau \ll 3t$, $D(t)$ reduces to $C(t)$ and one recovers Eq.\ (\ref{refaZZR}) which corresponds to static meroaxial
%disorder.

\begin{figure}
\subfigure{\resizebox{8cm}{!}
{\includegraphics{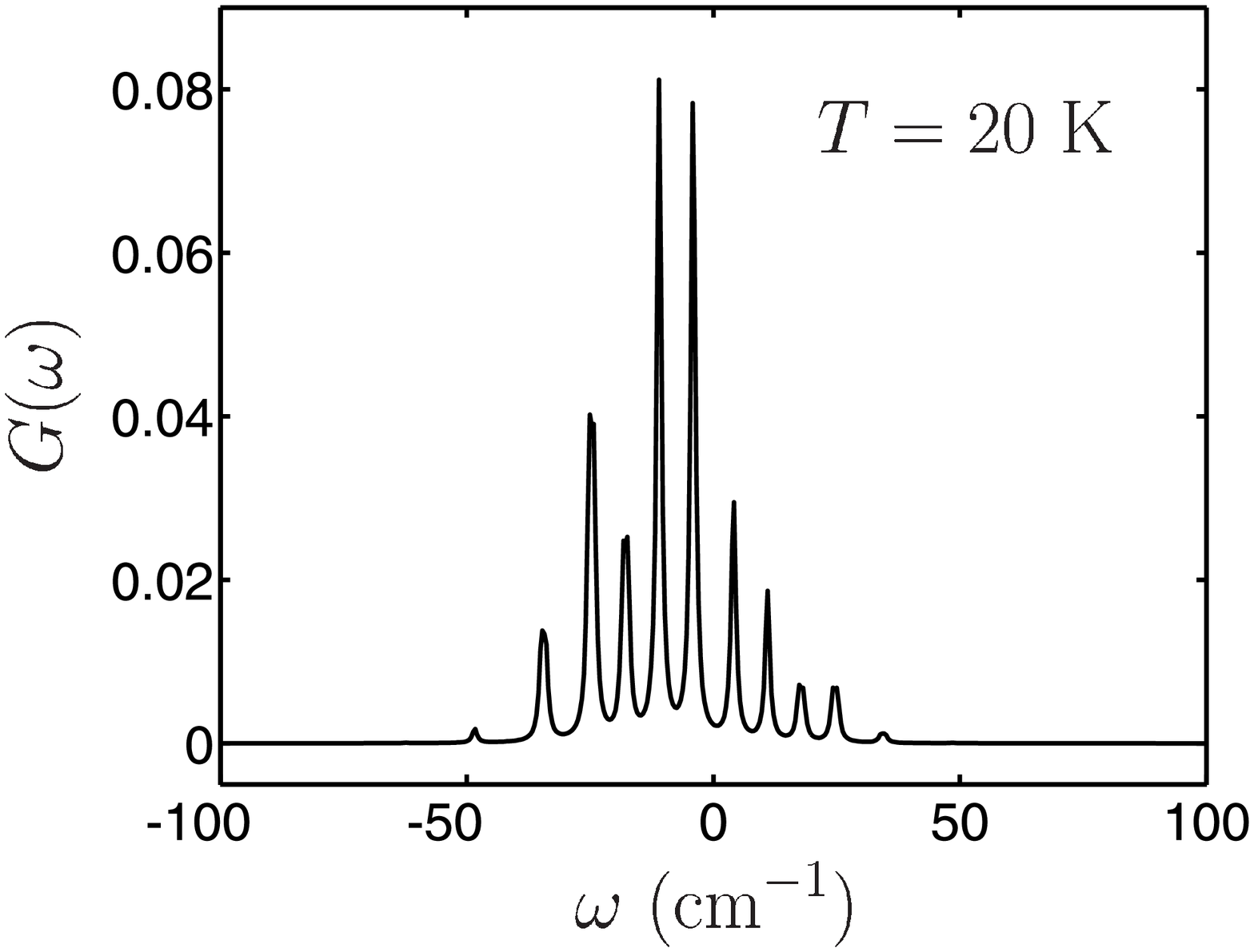}}} \\
\subfigure{\resizebox{8cm}{!}
{\includegraphics{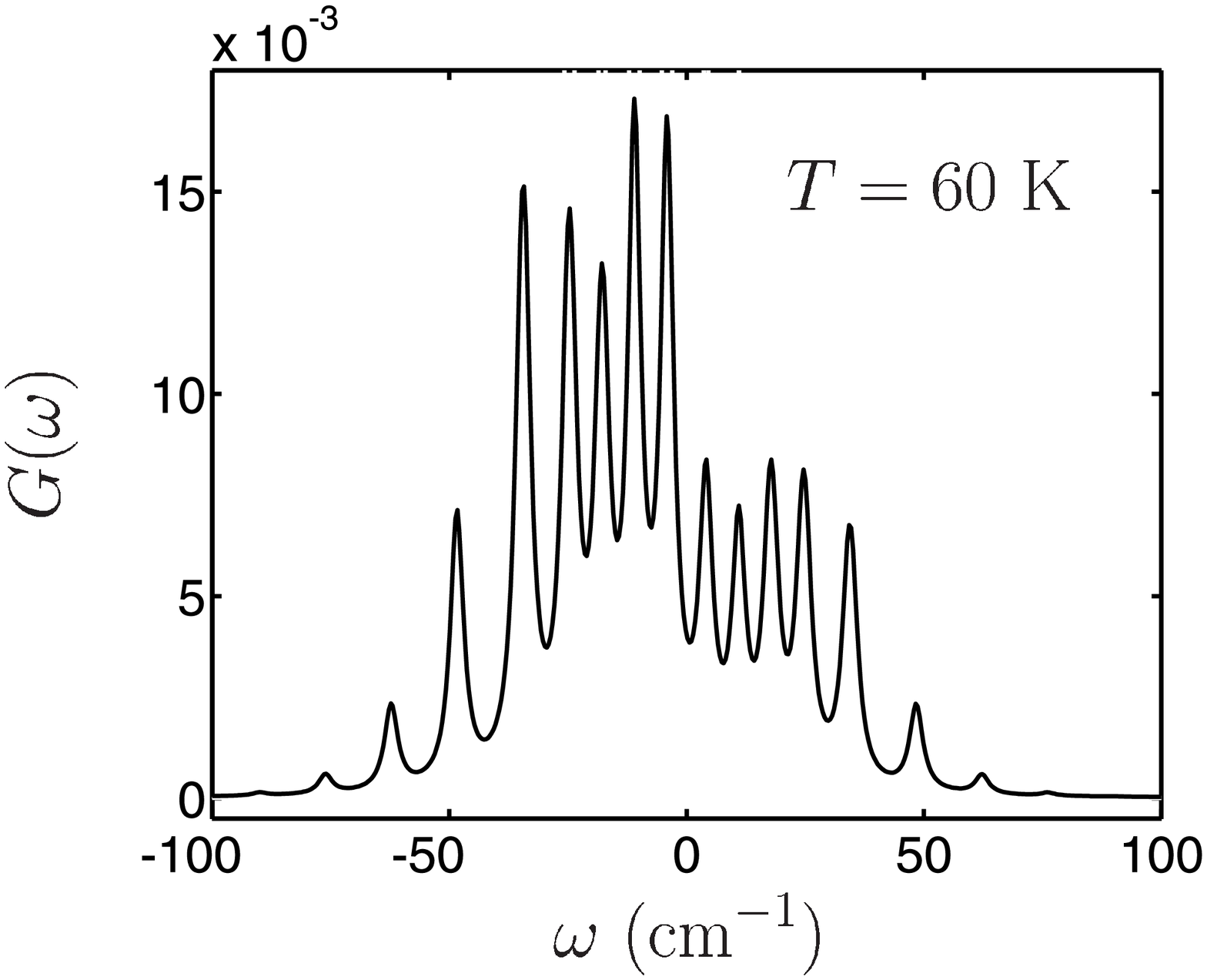}}} \\
\subfigure{\resizebox{8cm}{!}
{\includegraphics{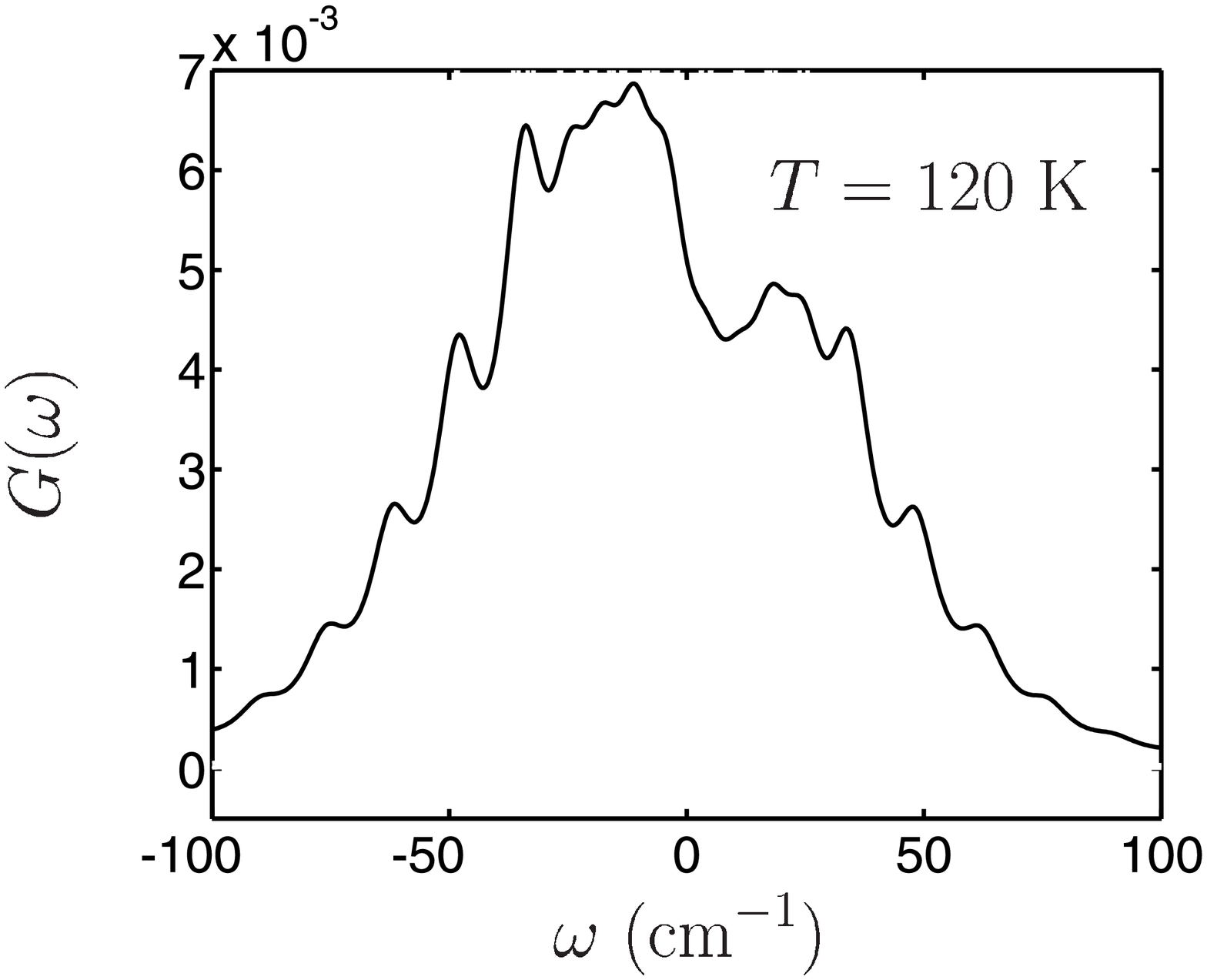}}}
\caption{Scattering function $G(\omega)$ for $T=20$ K, $60$ K and $120$ K.
}
\label{fig6}
\end{figure}

Similarly, using again Eq.\ (\ref{AtA0}) and the powder-averaged products $\overline{\bigl(A^{(i)}_{ZY}\bigr)^2}$ etc.\ in
Appendix \ref{appB}, we find
\begin{align}
   \overline{\bigl\langle A_{ZY}(t)A_{ZY}(0)\bigr\rangle} & = \frac{1}{15}e^{-3t/\tau}, \\
   \overline{\bigl\langle B_{ZY}(t)B_{ZY}(0)\bigr\rangle} & = \frac{11}{135}\left[1 + 2e^{-3t/\tau}\right], \\
   \overline{\bigl\langle C_{ZY}(t)C_{ZY}(0)\bigr\rangle} & = \frac{1}{135}\left[4 + 17e^{-3t/\tau}\right],
\end{align}
and hence
\begin{align}
   \overline{\bigl\langle \alpha_{ZY}^R(t)\alpha_{ZY}^R(0)\bigr\rangle} = \frac{b^2}{9}D(t),
\end{align}
with $D(t)$ again given by Eq.\ (\ref{Dt}).  In the limit of small relaxation time we recover Eq.\ (\ref{refaZYR}) for
static meroaxial disorder.  The Raman scattering laws are given by
\begin{align}
   \overline{R_{ZZZZ}(\omega)} = N\bigl(a^2\delta(\omega) + \frac{4}{27}b^2 D(\omega)\bigr), \label{f518}
\end{align}
and
\begin{align}
   \overline{R_{ZYZY}(\omega)} = N\frac{b^2}{9}D(\omega).
\end{align}
The Fourier transform of $D(t)$ leads to the scattering function
\begin{align}
   D(\omega) = C(\omega) + \frac{3}{5}J(\omega) + \frac{13}{5}G(\omega). \label{Domega}
\end{align}
The spectral function $C(\omega)$ is given by Eqs.\ (\ref{resComega}) -- (\ref{Lorentzian}) while
\begin{align}
   J(\omega) = \frac{1}{\pi}\frac{(3/\tau)}{\omega^2 + (3/\tau)^2}, \label{Jomega}
\end{align}
is the Fourier transform of the relaxation function $e^{-3t/\tau}$.  The Fourier transform of the interference
term
\begin{align}
   G(\omega) = \frac{1}{2\pi}\int_{-\infty}^{+\infty}dt\,e^{-i\omega t}C(t)e^{-3|t|/\tau}
\end{align}
is rewritten as
\begin{align}
   G(\omega) = \int_{-\infty}^{+\infty}d\omega'\,C(\omega - \omega')J(\omega'). \label{Gomega}
\end{align}
Using Eqs.\ (\ref{resComega}) -- (\ref{Lorentzian}) and (\ref{Jomega}) we obtain the scattering function
\begin{align}
   G(\omega) = G_{++}(\omega) + G_{--}(\omega) + G_{+-}(\omega) + G_{-+}(\omega).
\end{align}
The functions $G_{++}(\omega),\hdots,G_{--}(\omega)$ have the same structure as
$C_{++}(\omega),\hdots,C_{--}(\omega)$, Eqs.\ (\ref{Cpp}) -- (\ref{Cmp}), respectively, but where the Lorentzians
$F(\omega - \omega_{mm\pm 1}^{\sigma\sigma'})$, Eq.\ (\ref{Lorentzian}), are replaced by
\begin{align}
   H(\omega - \omega_{mm\pm 1}^{\sigma\sigma'}) = \frac{1}{2\pi}\left[\frac{\Gamma}
   {(\omega - \omega_{mm\pm 1}^{\sigma\sigma'})^2 + \Gamma^2}  \right].
\end{align}
Similarly to $C(\omega)$, Eq.\ (\ref{resComega}), the function $G(\omega)$ is a sum of weighted Lorentzians
centered around
$\omega = \omega_{mm\pm 1}^{\sigma\sigma'}$ but of width $2\Gamma$ where
\begin{align}
   \Gamma = 4D_\text{R} + 3/\tau. \label{gammaeq}
\end{align}
The broadening of
the transition frequencies of the quantum rotor with increasing temperature
is now due to the uniaxial rotational diffusion and the meroaxial
reorientations of the encapsulating C$_\text{84}$ molecule.  Notice that both contributions depend
on temperature.  In Fig.\ \ref{fig6} we have plotted the function $G(\omega)$ for several temperatures.  The
parameters describing the dynamics of the C$_\text{84}$ molecule are $\zeta_n=100$ for the rotational diffusion
model and $\tau_0^{-1} =
3\times 10^{12}$ s$^{-1}$ (attempt frequency) and $E_a = 580$ K (activation energy) for the thermally activated
meroaxial reorientations.
Comparable values of the activation energy, i.e.\ $32(7)$ meV and $35(15)$ meV have been deduced from neutron
scattering studies in solid C$_\text{70}$ \cite{Christides} and solid C$_\text{60}$ \cite{solidC60}, respectively.
  While for $T=20$ K and $60$
K the contribution of $3/\tau$ to the half width $\Gamma$ is negligible in comparison to $4D_\text{R}$, both
(additive) contributions become comparable at $150$ K.
At higher $T$ the thermally-activated reorientations are dominant
and lead to a smearing out of the low-frequency resonances
in the scattering function $G(\omega)$.

%  While the diffusion coefficient (see Eq.\ (\ref{diffcoeff}))

%---------
% Figure 6
%---------

%---------
% Figure 7
%---------
\begin{figure}
\subfigure{\resizebox{8cm}{!}
{\includegraphics{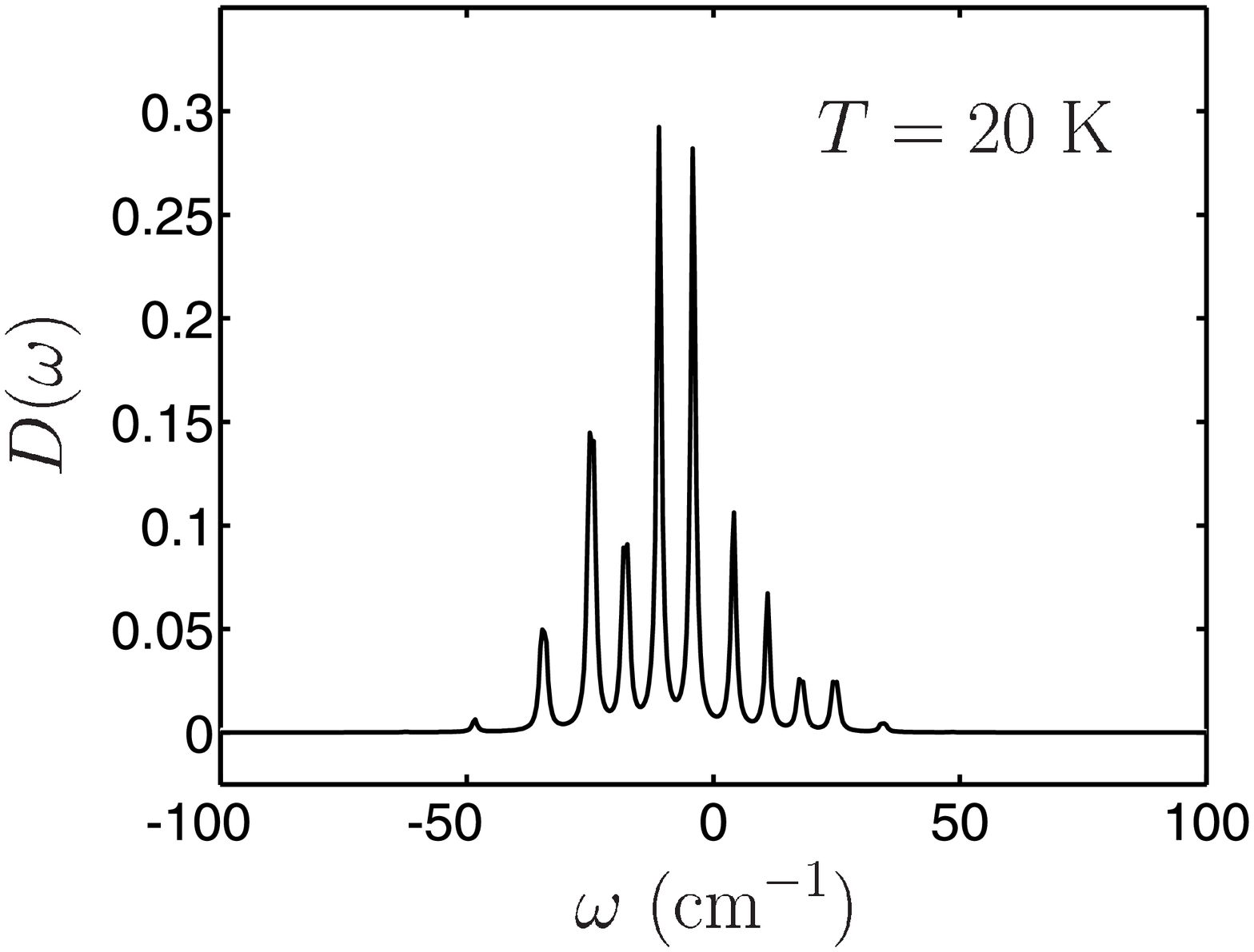}}} \\
\subfigure{\resizebox{8cm}{!}
{\includegraphics{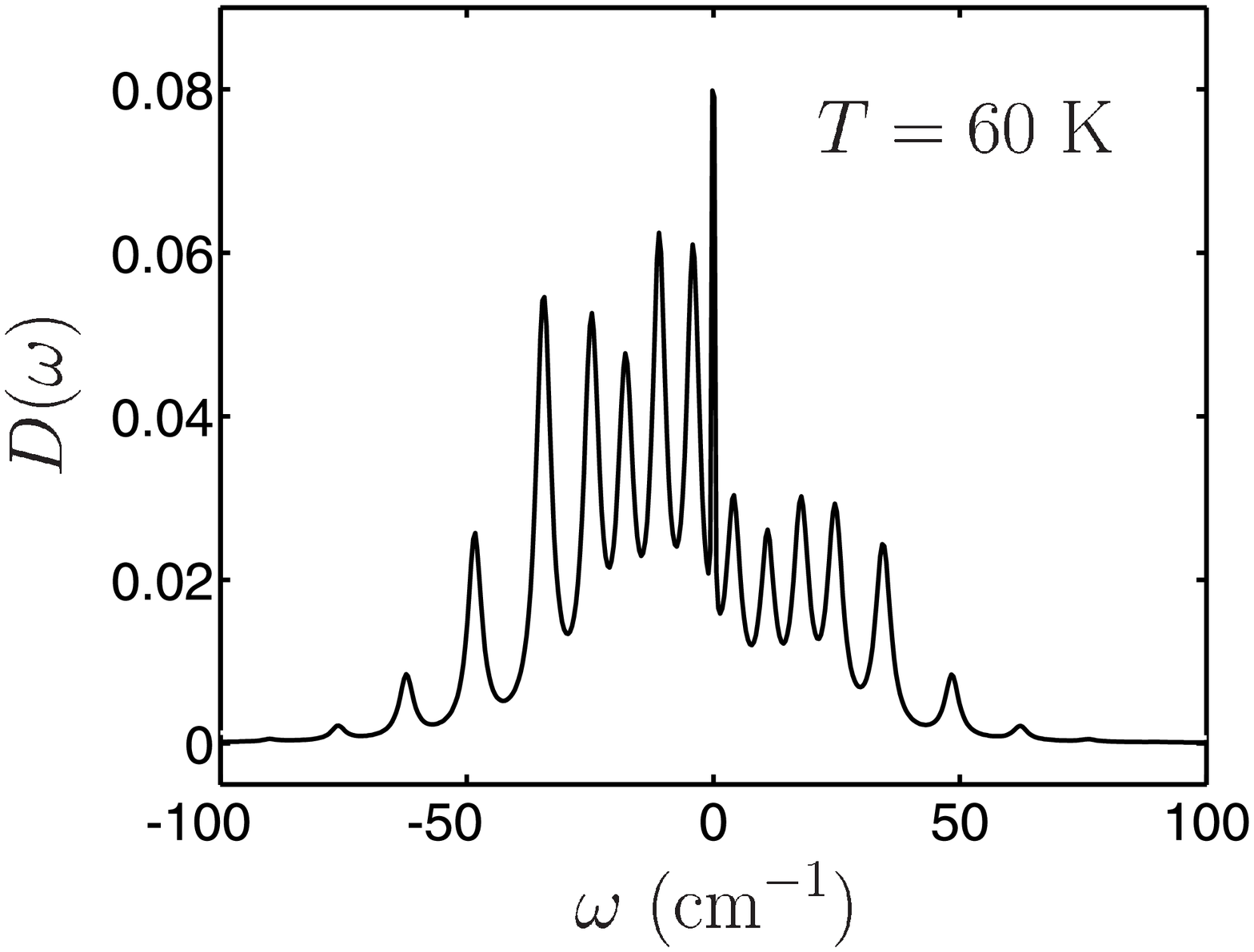}}} \\
\subfigure{\resizebox{8cm}{!}
{\includegraphics{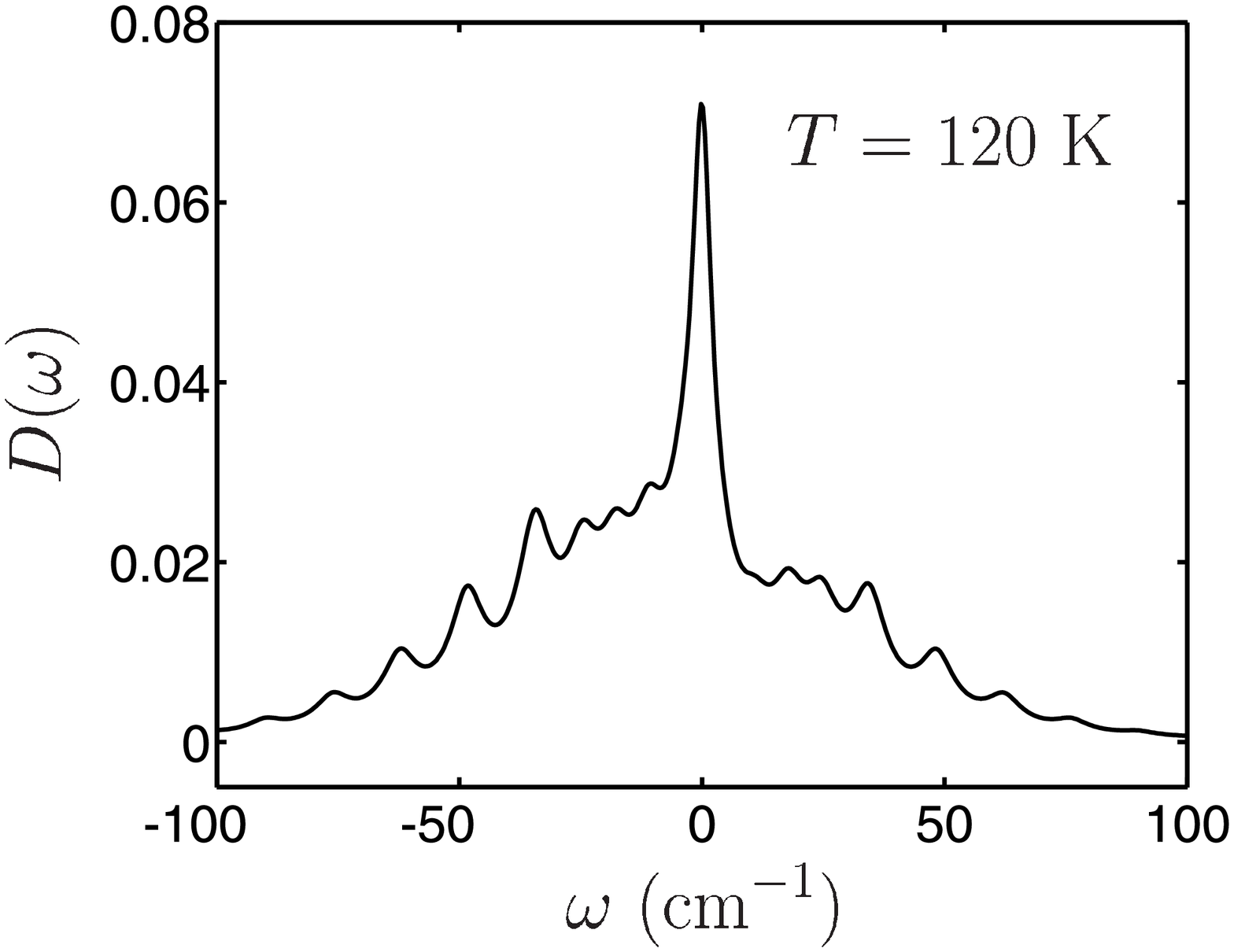}}}
\caption{Spectral function $D(\omega)$ of the low-frequency Raman scattering laws for $T=20$ K, $60$ K, $120$ K.
}
\label{fig7}
\end{figure}

If one would artificially exclude the tunneling motion, the function $D(\omega)$ entering the Raman scattering
laws of the C$_\text{2}$ unit would reduce to a superposition of Lorentzians centered at $\omega = 0$:
\begin{align}
   D(\omega) = F(\omega) + \frac{3}{5}J(\omega) + \frac{13}{5}H(\omega).
\end{align}
The first term on the right-hand side [given by Eq.\ (\ref{resFccomega})]  accounts solely for the rotational
uniaxial diffusion, the
second term for the meroaxial reorientations and the last term for the interference of these classical motions of
the encapsulating C$_\text{84}$ molecule.

%-------------------
\section{Discussion}
%-------------------
It has been shown that the low-frequency (rotational) part of the Raman scattering spectrum of a powder crystal of
Sc$_\text{2}$C$_\text{2}$@C$_\text{84}$ fullerite reflects the superposition of the quantum tunneling motion of
the encapsulated Sc$_\text{2}$C$_\text{2}$ complex about its long axis and the random
classical rotational motion of the
surrounding
C$_\text{84}$ molecule.
The effect of the C$_\text{84}$ molecule on the dynamics of Sc$_\text{2}$C$_\text{2}$ is twofold.
Firstly, since the long axis of Sc$_\text{2}$C$_\text{2}$ gyroscope
coincides with the $S_4$ axis of the
molecule, the rotation of Sc$_\text{2}$C$_\text{2}$ about this axis corresponds to the motion of the
C$_\text{2}$ bond as a planar quantum rotor in a fourfold potential \cite{6}.  Secondly,
any rotation
of the C$_\text{84}$ molecule caused by torques due to the thermal lattice environment
leads to a dragging of the enclosed Sc$_\text{2}$C$_\text{2}$ unit and
hence affects the spectrum of the C$_\text{2}$ quantum rotor seen in the laboratory frame.
The low-frequency Raman spectra resulting
from the interaction of the scattering radiation with the induced dipole of the C$_\text{2}$ rotor reflect these
features.

%We have developped a theory for the 
%low frequency (rotational) part of the Raman
%spectrum of a powder crystal of the endohedral fullerene Sc$_\text{2}$C$_\text{2}$@C$_\text{84}$.
%Since the long axis of the Sc$_\text{2}$C$_\text{2}$ complex
%coincides with the $S_4$ axis of the surrounding C$_\text{84}$ molecule, the
%rotation of the Sc$_\text{2}$C$_\text{2}$ complex about its long axis
%corresponds to the motion of C$_\text{2}$ as a planar quantum
%rotor in a fourfold potential.  We have shown that the spectrum which
%results from the interaction of the scattering radiation with the induced dipole of the C$_\text{2}$ unit
%reflects the
%superposition of two motions: the first one represents the quantum mechanical motion (tunneling) of the
%Sc$_\text{2}$C$_\text{2}$ rotor in the fourfold potential due to the encapsulating C$_\text{84}$ molecule while
%the second one represents the dragging of the Sc$_\text{2}$C$_\text{2}$ rotor by the classical motion of the
%C$_\text{84}$ molecule.

In analogy with the dynamics of the C$_\text{70}$ molecule in solid C$_\text{70}$ \cite{Christides}, we have
assumed that
the rotational motion of the C$_\text{84}$ molecule at a lattice site in a given crystallite
is composed of two parts:
uniaxial rotational diffusion about the $S_4$ axis and stochastic jumps of the $S_4$ orientation among $\langle 100
\rangle$ directions.
The superposition of the tunneling motion of the planar quantum rotor with the classical rotations of the
C$_\text{84}$ molecule leads to the spectral function
\begin{align}
   D(\omega) = C(\omega) + \frac{3}{5}J(\omega) + \frac{13}{5}G(\omega), \label{61}
\end{align}
given by Eq.\ (\ref{Domega}), in the Raman scattering laws $R_{ZZZZ}(\omega)$ and $R_{ZYZY}(\omega)$.

The function $C(\omega)$, defined by Eqs.\ (\ref{integral}) --
(\ref{Lorentzian}), accounts for tunneling transitions between the energy levels of the 
encapsulated C$_\text{2}$ rotor.
The spectrum consists of a series of resonances described by Lorentzians centered at the transition frequencies
(Table \ref{hbaromega}, $q=2$)
and broadened by the uniaxial rotational diffusion (half width $4D_\text{R}$) of the surrounding C$_\text{84}$
molecule.
Since the hindering potential for the rotational diffusion about the $S_4$ axis is weak, this motion affects the
spectrum already at low $T$.
  The temperature
dependence of the spectrum has been studied in Fig.\ \ref{Comegaplot}.

%The Raman spectrum measures transitions between the tunnel levels of
%C$_\text{2}$.  In a static fourfold potential, one would obtain perfectly sharp Raman lines.  In reality, the
%lines are broadened due to the motion of the C$_\text{84}$ molecule about its $S_4$ axis by a classical
%rotational diffusion model, characterized by a ``diffusion" constant $D_R$.  The Raman lines described by
%$C(\omega)$ correspond to the weighted Lorentzians, centered at the tunnel frequency transitions and of width
%$4D_R$ (see Fig. \ref{Comegaplot}).

The term $J(\omega)$ in Eq.\ (\ref{61}) accounts for the Raman spectrum of the radiation-induced
C$_\text{2}$ dipole while the Sc$_\text{2}$C$_\text{2}$
unit is dragged along by the classical reorientations of the C$_\text{84}$ molecule among its three meroaxial
directions.  This motion which reflects the changes of the orientation of the C$_\text{2}$ rotor plane is
described by a three sites stochastic jump model, characterized by a thermally
activated relaxation time $\tau = \tau_0e^{E_a/T}$.  Notice that $J(\omega)$ leads to a central resonance of half
width $(3/\tau)$ in the Raman scattering law even in absence of any quantum mechanical tunneling of
C$_\text{2}$.  The width of this central resonance (quasi-elastic peak) becomes appreciable at $T\ge 100$ K.  In
the scattering law $\overline{R_{ZZZZ}}(\omega)$, Eq.\ (\ref{f518}), this quasi-elastic peak is present in addition to the
elastic Rayleigh peak.  We suggest that in future low-energy Raman experiments additional attention will be given
to the possible identification of the temperature-dependent quasi-elastic peak.

The last term $G(\omega)$ in Eq.\ (\ref{61}) is due to the interference between the
uniaxial diffusion-modulated tunneling
motion described by $C(\omega)$ and the stochastic jump model accounted for by $J(\omega)$.  The function
$G(\omega)$ is a convolution of $C$ and $J$ [see Eq.\ (\ref{Gomega})].  While at low $T$ the spectra of
$C(\omega)$ and $G(\omega)$ are very similar (compare the plots for $T=20$ K, $T=60$ K in Fig.\ \ref{Comegaplot}
and Fig.\ \ref{fig6}) they become different at higher $T$ (see the $120$ K plots) where the increasing
influence of the stochastic jumps adds to the line broadening.
The width $2\Gamma$ of the individual resonances, Eq.\ (\ref{gammaeq}), increases from
$1.11$ cm$^{-1}$ at $T=20$ K to $3.37$ cm$^{-1}$ at $T=60$ K and $11.43$ cm$^{-1}$ at $T=120$ K.
This broadening leads to an overlap of the low-frequency resonances with increasing $T$.

Finally the sum $D(\omega)$ of these contributions which corresponds to the low-frequency Raman response function
is shown in Fig.\ \ref{fig7}.  The quasi-elastic peak centered at
$\omega = 0$ becomes important with increasing temperature.  In addition the
growing importance of $G(\omega)$ smears out the low-frequency resonances with increasing $T$ while the higher
frequency resonances remain prominent.

\begin{figure}
\subfigure{\resizebox{8cm}{!}
{\includegraphics{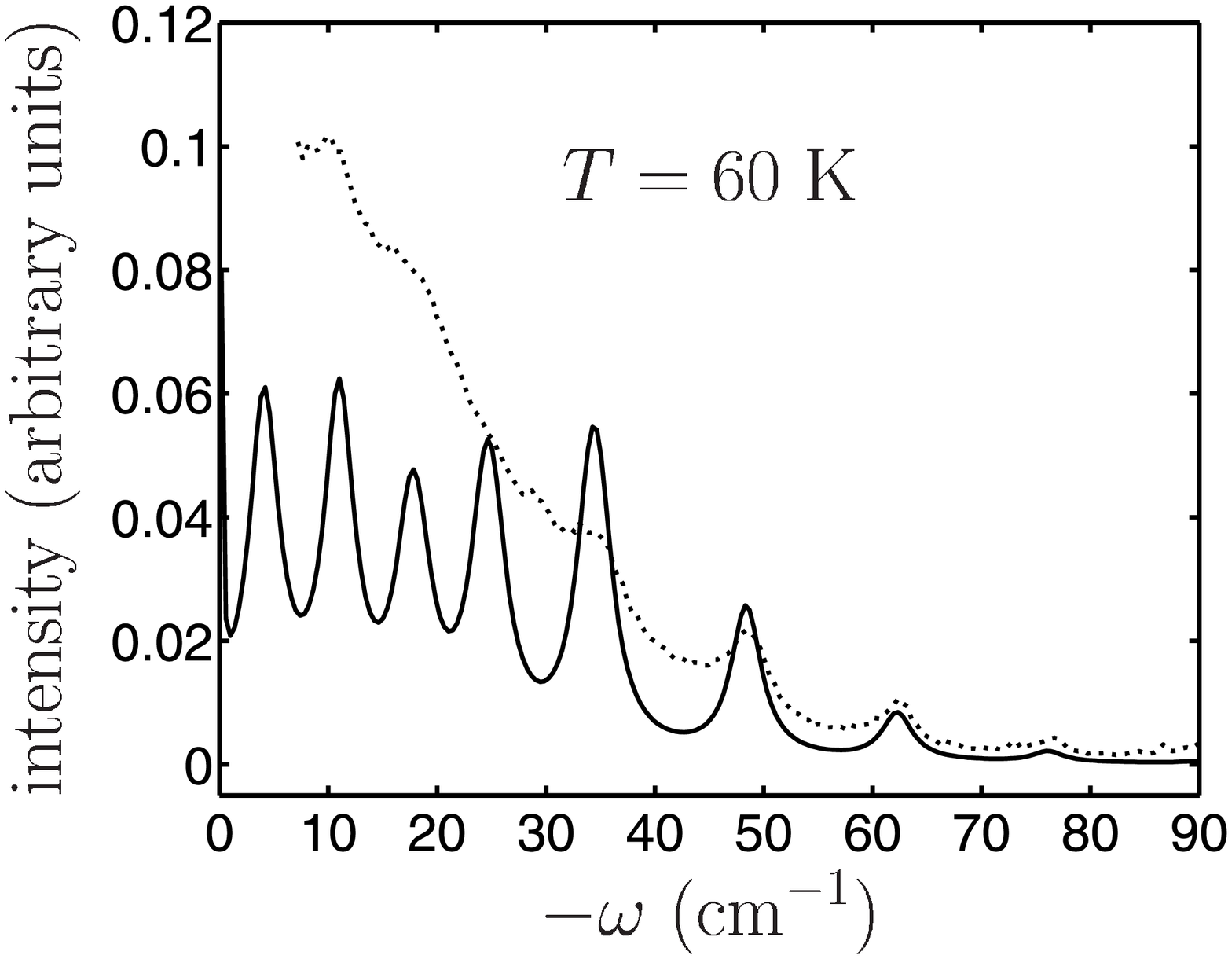}}} \\
\subfigure{\resizebox{8cm}{!}
{\includegraphics{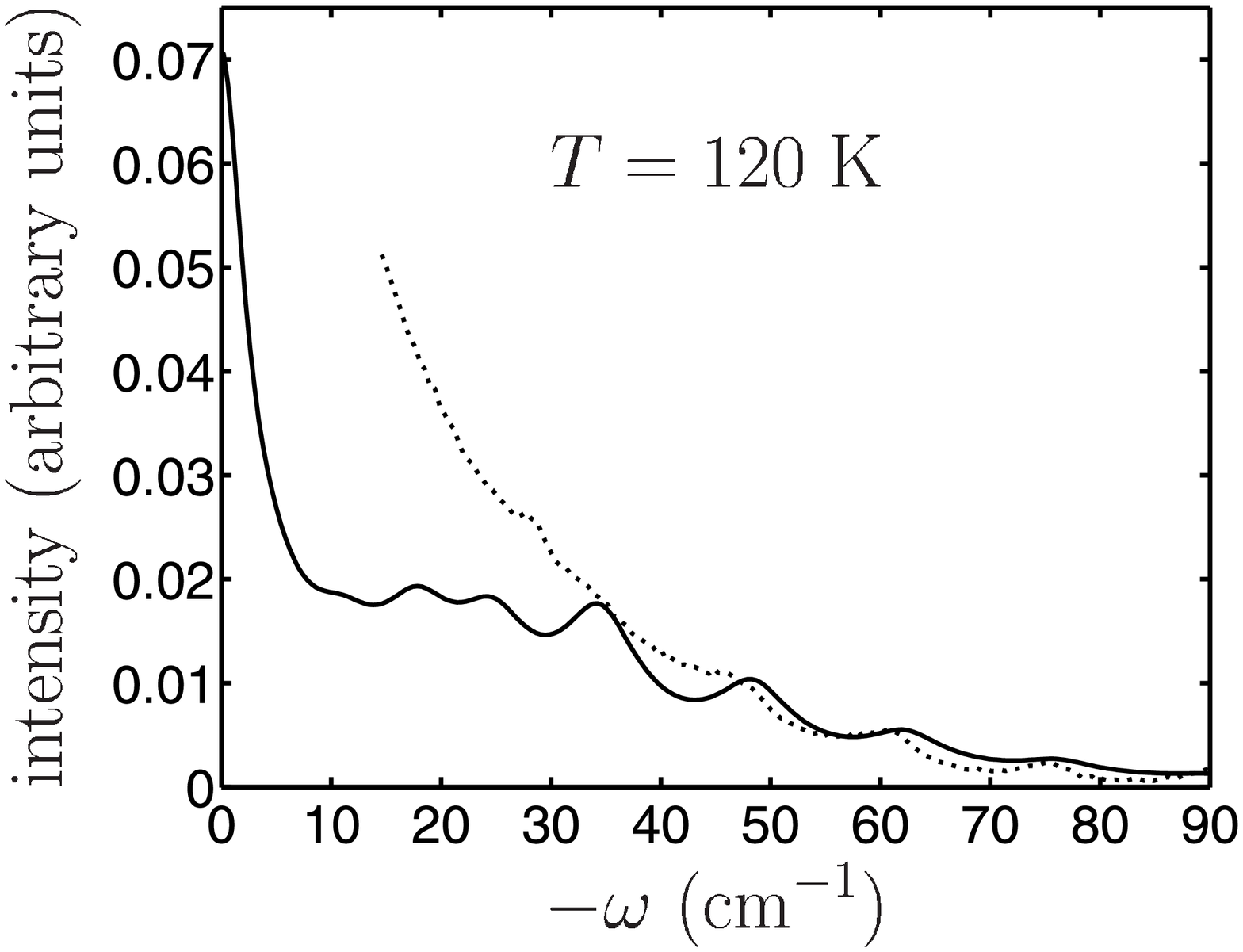}}} \\
\caption{Comparison of theoretical scattering law $D(\omega)$ (solid line),
calculated at $T = 60$ K and $T = 120$ K, with
experimental Raman results (dotted line) taken at the respective temperatures.}
\label{figexpthe}
\end{figure}

The overall shape of the spectral function $D(\omega)$ and its temperature evolution agree very well with the
low-frequency Raman scattering results of Ref.\ \onlinecite{6}.  There is quantitative agreement with the position
of the resonance lines.  The smearing out of the low-frequency resonances and the prominence of the
higher-frequency resonances with increasing $T$ (Fig.\ 3 of Ref.\ \onlinecite{6}) are well reproduced by the
present theory.  In addition to the positions of the resonance lines, the theory accounts for their
temperature-dependent broadening.
In Fig.\ \ref{figexpthe} we confront the theoretical spectra $D(\omega)$ with the experimental Raman spectra,
for both $T=60$ K and $T = 120$ K.
We notice that the experimental spectra are contaminated by a plasma line at $-\omega = 29.6$ cm$^{-1}$ \cite{6}.
Note that the central parts of the
experimental spectra have been
omitted in order to remove the effect of the unshifted Rayleigh peak.
On the other hand the theoretical spectrum
exhibits a quasi-elastic peak which is an intrinsic effect due to the
meroaxial stochastic reorientations of the Sc$_\text{2}$C$_\text{2}$@C$_\text{84}$ complex
[contribution $J(\omega)$ in $D(\omega)$].  Complementary to the present work it would be useful to
measure the
dynamics of the C$_\text{84}$ molecule in solid Sc$_\text{2}$C$_\text{2}$C$_\text{84}$ directly
say by NMR, neutron or $\mu$-spin spectroscopy.

%The contributions $J(\omega)$ and $G(\omega)$ in $D(\omega)$ are responsible
%for the frequencey-dependent ``background" (Fig.\ 2 of Ref.\ \onlinecite{6}) which tends to mask the low-frequency
%resonances with increasing $T$.

%---------
% Figure 8
%---------

\acknowledgments
The theoretical work has been supported by the Bijzonder Onderzoeksfonds, Universiteit Antwerpen (BOF-UA).
B.V.\ is a Postdoctoral Fellow of the Research Foundation - Flanders (FWO).
The experimental work has been supported by the EU Project NANOTEMP and by the Austrian FWF
(17345-PHY).

%\begin{acknowledgments}
%\end{acknowledgments}

%\pagebreak

\appendix
\section{}\label{appA}
We recall that $\nu$ is the angle of rotation of the C$_\text{84}$ molecule about its $S_4$ axis.  Treating the
molecule as a Brownian rotor we have the one-dimensional Langevin equation
\begin{align}
   \Theta\ddot{\nu}(t) + \zeta\dot{\nu}(t) = \lambda(t) - \frac{\partial U}{\partial\nu}.
\end{align}
Here $\Theta$ is the moment of inertia about the $S_4$ axis, $\zeta\dot{\nu}$ is the friction torque, $\lambda(t)$ a
white noise driving torque and $-\partial U/\partial \nu$ is a torque due to the orientational crystal field
potential $U\bigl(\nu(t)\bigr)$.  Under the assumption \cite{16} that the friction torque is dominant in comparison to the
inertial term and that the variation of $U$ with $\nu$ is weak, one can use standard methods \cite{11} to derive
a Smoluchowski equation for the angular distribution function $w\bigl(\nu(t)\bigr)$:
\begin{equation}
   \frac{\partial w}{\partial t} = D_\text{R}\frac{\partial^2 w}{\partial \nu^2}, \label{Smoluchowski}
\end{equation}
where $D_\text{R} = k_\text{B}T/\zeta$.

The functions
\begin{align}
   \bigl\{\psi_m(\nu)\bigr\} =
   \left\{\frac{e^{im\nu}}{\sqrt{2\pi}}\right\},
\end{align}
$m=0,\pm 1,\hdots$ are orthonormal eigenfunctions of the operator $\partial^2/\partial\nu^2$ with eigenvalues $\{0,-1,-m^2,\hdots\}$.
We calculate the conditional probability distribution $w(\nu,t|\nu_0,0)$ to find the molecule at an angle
$\nu$ at time $t$ when it was at angle $\nu_0$ at $t=0$.  The initial condition can be written as
\begin{align}
   \lim_{t\longrightarrow 0}w(\nu,t|\nu_0,0) = \delta(\nu-\nu_0) =
   \sum_m\psi_m^*(\nu)\psi_m(\nu_0), \label{wc0}
\end{align}
where the second member equality is just the closure relation.  On the other hand in the long-time limit the
orientation of the molecule should be random which corresponds to the condition 
\begin{align}
   \lim_{t\longrightarrow\infty} w(\nu,t|\nu_0,0) = \frac{1}{2\pi}. \label{wcinfinity}
\end{align}
A particular solution of (\ref{Smoluchowski}) subject to these boundary condition is of the form
\begin{align}
   w(\nu,t|\nu_0,0) = \sum_m e^{-D_\text{R}m^2t}\psi_m^*(\nu)\psi_m(\nu_0). \label{wc}
\end{align}

The correlation function $\bigl\langle\cos 2\nu(t)\cos 2\nu(0)\bigr\rangle$, Eq.\ (\ref{Fcct}), is rewritten as a
thermal average:
\begin{align}
   F^\text{cc}(t) & = \int_0^{2\pi}d\nu\,\int_0^{2\pi}d\nu_0\,\cos 2\nu G(\nu,t|\nu_0,0)\cos 2\nu_0. \label{A7}
\end{align}
Here $G_(\nu,t|\nu_0,0)d\nu d\nu_0$ is the joint probability of finding the C$_\text{84}$ molecule with
orientation angle $\nu_0\equiv \nu(0)$ in the interval $d\nu_0$ initially and orientation $\nu\equiv \nu(t)$ in $d\nu$
at time $t$.  One has
\begin{align}
   G(\nu,t|\nu_0,0) = w(\nu,t|\nu_0,0)p(\nu_0),
\end{align}
where the conditional probability $w(\nu,t|\nu_0,0)$ is given by
\begin{align}
   w(\nu,t|\nu_0,0) = \frac{1}{2\pi}\sum_me^{-m^2D_\text{R}t}e^{im(\nu-\nu_0)},
\end{align}
$m=0,\pm 1,\hdots$, and where
$p(\nu_0)=1/2\pi$ is the initial equilibrium probability.
%In the following we assume that the rotational diffusion coefficient $D_\text{R}$ is given by the Einstein
%relation
%\begin{align}
%   D_\text{R} = \frac{k_\text{B}T}{\zeta}, \label{diffcoeff}
%\end{align}
%where $\zeta$ is the friction coefficient and $T$ the temperature.
Carrying out the integrals in Eq.\ (\ref{A7}) gives as a result Eq.\ (\ref{Fcctb}).
In a similar way we obtain Eq.\ (\ref{Fsstb}).
%obtain
%\begin{align}
%   F^\text{cc}(t) = \bigl\langle\cos 2\nu(t) \cos 2\nu(0)\bigr\rangle = \frac{1}{2}e^{-4D_\text{R}t}. \label{Fcctb}
%\end{align}
%In a similar way we obtain for the correlation function $F^\text{ss}(t)$, Eq.\ (\ref{Fsst}),
%\begin{align}
%   F^\text{ss}(t) = \bigl\langle\sin 2\nu(t) \sin 2\nu(0)\bigr\rangle = \frac{1}{2}e^{-4D_\text{R}t}. \label{Fsstb}
%\end{align}

\section{}\label{appB}
Here we give details about the calculation of the powder averages in Sect.\ \ref{secPowder}.
%Finally the powder average is obtained by averaging the polarizability-polarizability
%correlation functions over the Euler angles.
We start from the situation where the C$_\text{84}$ molecule is in standard orientation, which corresponds to
the polarizabilities $\alpha_{ZZ}^{(1)}$ and $\alpha_{ZY}^{(1)}$ given by Eqs.\ (\ref{alphaZZ1}) and
(\ref{alphaZY1}) respectively.
In order to apply the rotation operation $R(\alpha,\beta,\gamma)$ we rewrite
the polarizabilities in terms of spherical harmonics $Y_l^m(\theta,\phi)$.  We use the notations and conventions
of Bradley and Cracknell \cite{BraCrack}.  With
\begin{align}
   Y_2^0(\theta) & = \left(\frac{5}{16\pi}\right)^{1/2}(3\cos^2\theta-1), \label{Y20} \\
   Y_2^{\pm 1}(\theta,\phi) & = \left(\frac{15}{8\pi}\right)^{1/2}\cos\theta\sin\theta e^{\pm i\phi} \label{Y21},
\end{align}
we get
\begin{align}
   \alpha_{ZZ}^{(1)} & = a + b\left(\frac{64\pi}{45}\right)^{1/2}\left.Y_2^0(\theta)\right|_{\phi = \pi/2}, \label{alphaZZ1a} \\
   \alpha_{ZY}^{(1)} & = -ib\left(\frac{8\pi}{15}\right)^{1/2}\left.\bigl[
   Y_2^1(\theta,\phi) - Y_2^{-1}(\theta,\phi)\bigr]\right|_{\phi = \pi/2}. \label{alphaZYa}
\end{align}
The condition $\phi = \pi/2$ ensures that the C$_\text{2}$ rotor initially moves in the $(Y,Z)$ plane.
Although the function $Y_2^0(\theta)$ does not depend on $\phi$, the condition $\phi = \pi/2$ has to be taken into
account after the application of a rotation operation.
The
transformation law of spherical harmonics under a rotation $R(\alpha,\beta,\gamma)$ reads:
\begin{align}
   R(\alpha,\beta,\gamma)Y_l^m(\theta,\phi) = \sum_{n=-l}^{+l}Y_l^n(\theta,\phi)
   {\EuScript D}_{nm}^l(\alpha,\beta,\gamma). \label{rotation}
\end{align}
Here ${\EuScript D}_{nm}^l(\alpha,\beta,\gamma)$ are the Wigner rotator functions defined by
\begin{align}
   {\EuScript D}_{nm}^l(\alpha,\beta,\gamma) = C_{nm}e^{-in\gamma}d_{nm}^l(\beta)e^{-im\alpha},
\end{align}
where $C_{nm} = i^{|n|+n}i^{-|m|-m}$ \cite{BraCrack}.  The functions $d_{nm}^l(\beta)$ are polynomials in $\sin(\beta/2)$ and
$\cos(\beta/2)$.
They satisfy the relations
\begin{align}
   d_{nm}^l(\beta) = (-1)^{n+m}d_{-n,-m}^l(\beta)=(-1)^{n+m}d_{mn}^l(\beta). \label{relations}
\end{align}
We recall that the angles $\theta$ and $\phi$ on the right-hand side of Eq.\ (\ref{rotation}) refer to the values before the
application of the rotation.  Applying  $R(\alpha,\beta,\gamma)$ to $\alpha_{ZZ}^{(1)}$, Eq.\ (\ref{alphaZZ1a}),
we have to evaluate
\begin{align}
   R(\alpha,\beta,\gamma)\left.Y_2^0(\theta)\right|_{\phi = \pi/2} = \sum_{n=-2}^{+2}Y_2^n(\theta,\pi/2){\EuScript D}_{n0}^2(\alpha,\beta,\gamma).
\end{align}
In addition to the definitions (\ref{Y20}) and (\ref{Y21}) we quote 
\begin{align}
   Y_2^{\pm 2}(\theta,\pi/2) = - \sqrt{\frac{15}{32\pi}}\sin^2\theta.
\end{align}
We further use
\begin{align}
   d_{20}^2(\beta) & = d_{-20}^2(\beta) = (\sqrt{6}/4)\sin^2\beta, \\
   d_{10}^2(\beta) & = d_{-10}^2(\beta) = -(\sqrt{6}/4)\sin 2\beta, \\
   d_{00}^2(\beta) & = (3\cos^2\beta - 1)/2.
\end{align}
We then find after some bookkeeping
\begin{align}
  R(\alpha,\beta,\gamma)\left.Y_2^0(\theta)\right|_{\phi = \pi/2} & = \left(\frac{45}{64\pi}\right)^{1/2}\bigl[A_{ZZ}^{(1)}(\beta,\gamma) +
  B_{ZZ}^{(1)}(\beta,\gamma)\sin 2\theta + C_{zz}^{(1)}(\beta,\gamma)\cos 2\theta \bigr],
\end{align}
where
\begin{align}
   A_{ZZ}^{(1)}(\beta,\gamma) & = \frac{1}{2}\left[\cos^2\beta - \sin^2\beta\cos 2\gamma - \frac{1}{3}\right], \\
   B_{ZZ}^{(1)}(\beta,\gamma) & = \sin 2\beta\sin\gamma, \\
   C_{ZZ}^{(1)}(\beta,\gamma) & = \frac{1}{2}\left[\sin^2\beta\cos 2\gamma +3\cos^2\beta - 1\right].
\end{align}
Hence
\begin{align}
   \alpha_{ZZ}^{(1)R} = a +
   b\bigl[A_{ZZ}^{(1)}(\beta,\gamma) + B_{ZZ}^{(1)}(\beta,\gamma)\sin 2\theta +
   C_{ZZ}^{(1)}(\beta,\gamma)\cos 2\theta\bigr]. \label{alphaZZR}
\end{align}
With the change $(\alpha,\beta,\gamma)\longrightarrow (\alpha,\beta,\gamma-\frac{\pi}{2})$ we find the
coefficients entering $\alpha_{ZZ}^{(2)R}$
\begin{align}
   A_{ZZ}^{(2)}(\beta,\gamma) & = \frac{1}{2}\left[\cos^2\beta + \sin^2\beta\cos 2\gamma - \frac{1}{3}\right], \\
   B_{ZZ}^{(2)}(\beta,\gamma) & = -\sin 2\beta\sin\gamma, \\
   C_{ZZ}^{(2)}(\beta,\gamma) & = \frac{1}{2}\left[-\sin^2\beta\cos 2\gamma +3\cos^2\beta - 1\right],
\end{align}
and with $(\alpha,\beta,\gamma)\longrightarrow (\alpha,\beta+\frac{\pi}{2},\gamma=0)$ the
coefficients entering $\alpha_{ZZ}^{(3)R}$
\begin{align}
   A_{ZZ}^{(3)}(\beta,\gamma) & = \frac{1}{2}\left[\sin^2\beta - \cos^2\beta - \frac{1}{3}\right], \\
   B_{ZZ}^{(3)}(\beta,\gamma) & = 0, \\
   C_{ZZ}^{(3)}(\beta,\gamma) & = \frac{1}{2}\left[\cos^2\beta +3\sin^2\beta - 1\right].
\end{align}
The average polarizability $\alpha_{ZZ}^R$ is then given by Eq.\ (\ref{alphaZZRmeroaxial}).
Applying the definition of powder average Eq.\ (\ref{Fbetagammaaverage}) we get
$\overline{[A_{ZZ}^{(i)}]^2} = \frac{4}{45}$, $i=1,2,3$;
$\overline{[B_{ZZ}^{(i)}]^2} = \frac{4}{15}$, $i=1,2$;
$\overline{[C_{ZZ}^{(i)}]^2} = \frac{4}{15}$, $i=1,2$; $\overline{[C_{ZZ}^{(3)}]^2} = \frac{8}{15}$; 
$\overline{[A_{ZZ}^{(i)}A_{ZZ}^{(j)}]^2} = -\frac{2}{45}$, $i\ne j$;
$\overline{[B_{ZZ}^{(i)}B_{ZZ}^{(j)}]^2} = 0$, $i\ne j$;
$\overline{[C_{ZZ}^{(i)}C_{ZZ}^{(j)}]^2} = -\frac{2}{15}$, $i=1,2$, $j=3$;
$\overline{[C_{ZZ}^{(1)}C_{ZZ}^{(2)}]^2} = \frac{2}{15}$.

In a similar way we apply the rotation operation to $\alpha_{ZY}^{(1)}$, Eq.\ (\ref{alphaZYa}), and calculate 
\begin{align}
   R(\alpha,\beta,\gamma)Y_2^{\pm 1}(\theta,\pi/2) = \sum_{n=-2}^{+2}Y_2^n(\theta,\pi/2){\EuScript
   D}_{n1}^2(\alpha,\beta,\gamma).
\end{align}
Making use of
\begin{xalignat}{2}
   d_{21}^2(\beta) & = -\cos^2\frac{\beta}{2}\sin\beta, & & d_{11}^2(\beta) = \cos^2\frac{\beta}{2}(2\cos\beta - 1),
   \\
   d_{-21}^2(\beta) & = \sin^2\frac{\beta}{2}\sin\beta, & & d_{-11}^2(\beta) = \sin^2\frac{\beta}{2}(2\cos\beta +
   1), \\
   d_{01}^2(\beta) & = \sqrt{6}/4\sin 2\beta &&
\end{xalignat}
and of the relations (\ref{relations}) we find
\begin{multline}
   (-i)R(\alpha,\beta,\gamma)\left.\bigl[Y_2^1(\theta,\phi) - Y_2^{-1}(\theta,\phi)\bigr]\right|_{\phi = \pi/2}
    \\ = \left(\frac{15}{8\pi}\right)^{1/2}\bigl[A_{ZY}^{(1)}(\alpha,\beta,\gamma) + B_{ZY}^{(1)}(\alpha,\beta,\gamma)\sin 2\theta +
      C_{ZY}^{(1)}(\alpha,\beta,\gamma)\cos 2\theta\bigr]
\end{multline}
where
\begin{align}
   A_{ZY}^{(1)}(\alpha,\beta,\gamma) & = \frac{1}{2}\left[
     \sin\beta\sin2\gamma\cos\alpha + \frac{\sin2\beta\cos 2\gamma\sin\alpha}{2}
      + \frac{\sin 2\beta\sin\alpha}{2}
   \right], \\
   B_{ZY}^{(1)}(\alpha,\beta,\gamma) & = \cos\beta\cos\gamma\cos\alpha - \cos 2\beta\sin\gamma\sin\alpha, \\
   C_{ZY}^{(1)}(\alpha,\beta,\gamma) & = \frac{1}{2}\left[-\sin\beta\sin2\gamma\cos\alpha - 
   \frac{\sin 2\beta\cos 2\gamma\sin\alpha}{2} + \frac{3\sin 2\beta\sin\alpha}{2}\right].
\end{align}
Hence
\begin{align}
  \alpha_{ZY}^{(1)R} = b\bigl[A_{ZY}^{(1)}(\alpha,\beta,\gamma) + B_{ZY}^{(1)}(\alpha,\beta,\gamma)
       + C_{ZY}^{(1)}(\alpha,\beta,\gamma)  \bigr].
\end{align}

With the change of angles $(\alpha,\beta,\gamma - \frac{\pi}{2})$ we get the coefficients entering
$\alpha_{ZY}^{(2)R}$:
\begin{align}
   A_{ZY}^{(2)}(\alpha,\beta,\gamma) & = \frac{1}{2}\left[
     -\sin\beta\sin2\gamma\cos\alpha - \frac{\sin 2\beta\cos 2\gamma\sin\alpha}{2}
      + \frac{\sin 2\beta\sin\alpha}{2}
   \right], \\
   B_{ZY}^{(2)}(\alpha,\beta,\gamma) & = \cos\beta\sin\gamma\cos\alpha + \cos 2\beta\cos\gamma\sin\alpha, \\
   C_{ZY}^{(2)}(\alpha,\beta,\gamma) & = \frac{1}{2}\left[-\sin\beta\sin 2\gamma\cos\alpha + 
   \frac{\sin 2\beta\cos 2\gamma\sin\alpha}{2} + \frac{3\sin 2\beta\sin\alpha}{2}\right],
\end{align}
and with $(\alpha,\beta,\gamma)\longrightarrow (\alpha,\beta+\frac{\pi}{2},\gamma = 0)$ the coefficients entering 
$\alpha_{ZY}^{(3)R}$:
\begin{align}
   A_{ZY}^{(3)}(\alpha,\beta,\gamma) & = -\frac{1}{2}\sin 2\beta\sin\alpha, \\
   B_{ZY}^{(3)}(\alpha,\beta,\gamma) & = -\sin\beta\cos\alpha, \\
   C_{ZY}^{(3)}(\alpha,\beta,\gamma) & = -\frac{1}{2}\sin 2\beta\sin \alpha.
\end{align}
With the definition of powder average Eq.\ (\ref{Falphabetagammaaverage}) we get 
$\overline{[A_{ZY}^{(i)}]^2} = \frac{1}{15}$, $i=1,2,3$;
$\overline{[B_{ZY}^{(i)}]^2} = \frac{1}{5}$, $i=1,2$;
$\overline{[B_{ZY}^{(3)}]^2} = \frac{1}{3}$;
$\overline{[C_{ZY}^{(i)}]^2} = \frac{1}{5}$, $i=1,2$;
$\overline{[C_{ZY}^{(3)}]^2} = \frac{1}{15}$;
$\overline{A_{ZY}^{(i)}A_{ZY}^{(j)}} = -\frac{1}{30}$, $i\ne j$;
$\overline{B_{ZY}^{(i)}B_{ZY}^{(j)}} = 0$, $i\ne j$;
$\overline{C_{ZY}^{(i)}C_{ZY}^{(3)}} = -\frac{1}{10}$, $i=1,2$;
$\overline{C_{ZY}^{(1)}C_{ZY}^{(2)}} = \frac{1}{10}$.

%-----------
% References
%-----------

\end{document}